\documentclass[10pt, a4paper]{article}
\usepackage{graphicx}
\usepackage{hyperref}
\usepackage{longtable}
\usepackage{booktabs}
\usepackage{boxedminipage}
\usepackage{chngcntr}
\usepackage{upquote}
\usepackage{etoolbox}
\patchcmd{\thebibliography}{\section*{\refname}}{}{}{}

\title{Healthcare serial killer or coincidence? \\
Statistical issues in investigation of suspected medical misconduct}
\author{P.J. Green\footnote{University of Bristol,UK}, R.D. Gill\footnote{Leiden University, Netherlands; email: {\tt gill@math.leidenuniv.nl}}, N. Mackenzie\footnote{Arnot Manderson Advocates, Edinburgh}, J. Mortera\footnote{Universit\'a Roma Tre}, W.C. Thompson\footnote{University of California, Irvine}}
\date{27 September, 2022}

\begin{document}
\maketitle

\begin{abstract}
Justice systems are sometimes called upon to evaluate cases in which
healthcare professionals are suspected of killing their patients
illegally. These cases are difficult to evaluate because they involve at
least two levels of uncertainty. Commonly in a murder case it is clear
that a homicide has occurred, and investigators must resolve uncertainty
about who is responsible. In the cases we examine here there is also
uncertainty about whether homicide has occurred. Investigators need to
consider whether the deaths that prompted the investigation could
plausibly have occurred for reasons other than homicide, in addition to
considering whether, if homicide was indeed the cause, the person under
suspicion is responsible. In this report (commissioned by the Section on Forensic Statistics
of the Royal Statistical Society, London) we provide advice and guidance on the
investigation and evaluation of such cases. Our work was prompted by
concerns about the statistical challenges such cases pose for the legal
system.
\end{abstract}

%\section{Table of Contents}
\tableofcontents

{
\counterwithin*{subsection}{section}
\renewcommand{\thesubsection}{\thesection.\alph{subsection}}

\section{Overview}

Justice systems are sometimes called upon to evaluate cases in which
healthcare professionals are suspected of killing their patients
illegally. These cases are difficult to evaluate because they involve at
least two levels of uncertainty. Commonly in a murder case it is clear
that a homicide has occurred, and investigators must resolve uncertainty
about who is responsible. In the cases we examine here there is also
uncertainty about whether homicide has occurred. Investigators need to
consider whether the deaths that prompted the investigation could
plausibly have occurred for reasons other than homicide, in addition to
considering whether, if homicide was indeed the cause, the person under
suspicion is responsible.

In this report, the RSS provides advice and guidance on the
investigation and evaluation of such cases. This report was prompted by
concerns about the statistical challenges such cases pose for the legal
system. The cases often turn, in part, on statistical evidence that is
difficult for lay people and even legal professionals to evaluate.
Furthermore, the statistical evidence may be distorted by biases, hidden
or apparent, in the investigative process that render it misleading. In
providing advice on how to conduct investigations in such cases, this
report particularly focuses on minimising the kinds of biases that could
distort statistical evidence arising from the investigation. This report
also provides guidance on how to recognise and take account of such
biases when evaluating statistical evidence and more broadly on how to
understand the strengths and limitations of such evidence and give it
proper weight.

This report is designed specifically to help all professionals involved
in investigating such cases and those who evaluate such cases in the
legal system, including expert witnesses. It will also be of interest to
scholars and legal professionals who are interested in the role of
statistics in evidentiary proof, and more generally to anyone interested
in improving criminal investigations. With such a wide range of
audiences, it is inevitable that for some readers certain sections may
seem more relevant, and some less so, but we believe it is important not
to aim particular sections at particular kinds of reader. We want, for
example, the barrister to see what advice we give to the expert
statistical witness -- and we hope understand it, at least in broad
terms -- and vice versa; we believe that is important in helping all
parties to appreciate the contributions of others in reaching just
outcomes.

Because suspicions about medical murder often arise due to a surprising
or unexpected series of events, such as an unusual number of deaths
among patients under the care of a particular professional, this report
will begin (in \textbf{Section 2}) with a discussion of the statistical
challenge of distinguishing event clusters that arise from criminal acts
from those that arise coincidentally from other causes. This analysis
will show that seemingly improbable patterns of events (eg apparent
clusters, rising trends, etc.) can often arise without criminal
behaviour and may therefore have less probative value than people assume
for distinguishing criminality from coincidence.

\textbf{Section 3} of this report will focus on the competing theories
that are often advanced by the prosecution and defence when a medical
professional faces criminal charges for killing patients. The
prosecution's theory is typically that a medical professional,
previously trusted to perform critical life-saving functions, has
unexpectedly (and sometimes inexplicably), chosen to murder patients in
his or her care. While history has shown that humans are capable of such
behaviour, and there have indeed been cases in which, for example,
physicians have murdered multiple patients, nevertheless proven
instances are thankfully extraordinarily rare -- a mere handful of
documented cases, perhaps a dozen or so per year, among the many
millions of healthcare professionals worldwide. So the prosecution's
theory in such cases is often one that appears, \emph{a priori}, to be
improbable. Alternative theories -- ie, that some unknown factors, or
mere chance, caused deaths to occur in apparently extraordinary numbers
among patients under the care of a particular professional -- often also
appear improbable. So the assessment of the case invariably turns, at
least in part, on a weighing or balancing of the probabilities of
seemingly extraordinary events. Such assessments are challenging under
the best of circumstances but become especially difficult when the
evidence adduced to distinguish between the competing theories may be
biased or presented in a misleading manner.

\textbf{Section 4} of this report discusses the kinds of investigative
biases that can arise in these cases. Our focus is on ways that
investigators' desires and expectations may
\emph{\textbf{unintentionally and even unconsciously}} influence what
they look for, how they characterise and classify what they find, what
they deem to be relevant and irrelevant, and what they choose to
disclose. Examiner bias is a well-known phenomenon in both scientific
and forensic investigations. It arises in large part from what are known
as observer effects, a tendency for human beings to look for data
confirming their expectations (confirmation bias) and to interpret data
in ways that are subtly (and often unconsciously) influenced by their
expectations and desires. Statisticians have long studied the ways in
which examiner bias can distort statistical evidence emerging from
scientific and forensic investigations. In Section 4, we apply insights
from this scientific literature to an analysis of the investigative
process in the types of cases discussed in this report. We also draw
examples from investigations of actual cases that illustrate what we
believe to have been biased investigative processes and discuss how such
biases can generate misleading statistical findings. It bears repeating
that our focus in this section is on processes that can
\emph{unintentionally and unconsciously} influence the investigative
process. We are not questioning the general honesty, integrity or good
intentions of those involved in investigating such cases. We focus
instead on investigative procedures that can distort statistical
findings in ways that, while entirely unintentional, may nevertheless be
important.

\textbf{Section 5} of this report provides advice on how to improve
investigative procedures in order to minimise investigative biases.
While it is impossible to eliminate all human biases from a criminal
investigation, there are a number of procedures that can reduce bias and
thereby improve the quality and objectivity of the evidence emerging
from the types of investigations we discuss here. We focus particularly
on the advantages of blinding and masking procedures, which involve
temporarily withholding potentially biasing facts from some of those
involved in the investigation. We go on to discuss ways to reduce
``tunnel vision'' in which the investigation becomes a search for
evidence confirming a particular investigative theory while ignoring or
dismissing evidence inconsistent with that theory. We provide and
explain advice on appropriate correct analyses of data, and discuss two
worked examples.

\textbf{Section 6} provides advice on evidence evaluation and
fact-finding in these cases. We expect this report to be relevant and
useful anywhere such cases may arise; hence we do not limit our
discussion to the needs of a particular legal system, and expect our
advice to be useful both in inquisitorial and adversarial legal
processes. We believe the statistical issues in these cases pose
challenges to legal fact-finders in every jurisdiction, whether they are
professional judges or lay jurors, and are challenging for lawyers as
well. Our advice focuses on identifying and appreciating ways in which
statistical evidence may be misleading, and assuring (to the extent
possible) that presentations of evidence are balanced in order to help
triers-of-fact appreciate both the strengths and limitations of the
evidence, and give it only the weight it deserves. We will provide
examples of presentations and arguments that we consider to be
misleading or inappropriate. We will discuss cautionary instructions
that may be helpful to lay fact-finders. Ultimately, we hope our
comments will help lawyers and judges, and statisticians and other
experts, refine their presentation and evaluation of evidence in these
difficult cases in order to better serve the interests of justice.

\textbf{Finally}, we draw together our main conclusions, and present a
summary of our most important recommendations in Section 7.

\section{``This could not have been a coincidence!''}

\textbf{The challenge of drawing
conclusions from suspicious clusters of deaths (or other adverse
outcomes)}

In some cases suspicions against medical professionals arise for the
very reason that an apparently unusual number of deaths occurs among
their patients. In other cases suspicions arise for unrelated reasons
and this prompts an examination of cases where a certain medical
professional was on duty and this reveals an apparently unusual number
of deaths. There is a statistical challenge of distinguishing event
clusters that arise from criminal acts from those that arise
coincidentally from other factors. Seemingly improbable clusters of
events can often arise by chance without criminal behaviour and may
therefore have less probative value than people assume for
distinguishing criminality from coincidence.

Lucy and Aitken published an analysis of evidence used to prosecute
medical professionals accused of harming their patients\footnote{Lucy
  and Aitken (2002).}. They found (see p. 152) that ``evidence of
attendance'' was ``by far the most frequently occurring'' yet was also
``the most difficult type of evidence, both from a legal and
epistemological point of view''. While other types of evidence may be
presented, such as evidence of a criminal intent (\emph{mens rea}) and
the means to carry it out, or eyewitness accounts, these tend to be less
than definitive for a variety of reasons, such as the difficulty of
ascertaining retrospectively the exact cause of death and the
uncertainty inherent in assessing human motives and behaviour.
Statistics on the relative rate of deaths when a particular professional
was ``in attendance'' may, by contrast, seem more objective and
scientific, making statistical evidence the lynchpin of these cases.

Drawing causal conclusions from a statistically improbable cluster of
events is often challenging, however.\footnote{Wartenberg, 2001.} A
criminal investigation is analogous to a retrospective observational
study. In such a study, it is possible to ascertain correlations between
variables. The study might establish, for example, that the death rate
was higher when a particular medical professional was present on a
hospital ward. However, one of the fundamental principles of logical
inference is that correlation does not prove causation. The increase in
death rate cannot, in itself, prove that the professional in question
was engaging in misconduct that caused the increase in deaths because
other factors, known as confounding variables, might offer alternative
explanations.\footnote{A confounder is a variable, not of prime concern
  in a study, that is associated with both the `exposure' (eg presence
  of a particular nurse) and the `outcome' (eg unexpected death of a
  patient). Neglect or inadequate attention to confounders typically
  leads to misleading conclusions about the causal effect of the
  exposure.} Competent investigators are attentive to the possibility of
confounding variables and may attempt to take them into account. Even if
all known confounding variables are taken into account, however, there
might be additional confounders, unknown, unmeasured, unmeasurable, or
otherwise inadequately dealt with, that affect mortality rates when a
given medical professional is on duty. For example, there may be changes
in the circumstances and characteristics of the hospital for reasons
that are not measured, or even not observable at all. So finding an
association of a particular professional with high mortality rates
cannot \emph{per se} have a causal interpretation.

It is customary to compute the relative risk, which is the ratio of the
death rate per unit of time when the suspected individual is on duty to
the death rate per unit time when the suspect is not on duty. For
example, an analysis in a prosecution against US nurse Jane Bolding (see
box over page) found that patients under her care were 47.5 times more
likely to suffer cardiac arrest than were those of other nurses, and
that ``an epidemic'' of cardiac arrests ceased when Bolding left the
hospital unit where it occurred (\emph{Sacks} et al., 1988).

To help the court interpret such statistics, experts often report a
\emph{p}-value, which is an estimate of the probability that the
observed (or a greater) number of deaths would occur by chance, if the
risk to the patients in question was in fact no higher than the risk to
similar patients who were not under the care of the accused.

\bigskip

\begin{boxedminipage}{0.95\textwidth}

\emph{\textbf{The case of Jane Bolding}}

\medskip

\emph{Statistical evidence often plays a prominent role in the
investigation of suspected healthcare serial killers. In 1988, Jane
Bolding, an American nurse who worked in the intensive care unit of
Prince George's Medical Center in Maryland, was prosecuted for
serial murder of patients, allegedly by administering lethal doses of
potassium chloride. The key evidence against Bolding was the high
incidence of cardiac arrest during periods when Bolding was on duty.
Evidence suggested that she had been the primary nurse on duty when 57
heart attacks occurred, while the number during comparable periods when
other nurses were on duty had never exceeded five. An analysis by
epidemiologists from the U.S. Centers for Disease Control (CDC)
concluded that Bolding's patients were 47.5 times more likely to
experience cardiac arrest than those of other nurses and that``an
epidemic'' of cardiac arrests ceased when Bolding left the hospital unit
where it occurred (Sacks et al., 1988; CDC, 1985). Sacks testified at
Bolding's trial that ``[t]he chances of [this large
number of cardiac arrests] happening by chance is about one in 100
trillion.''}

\medskip

\emph{Other than the statistical evidence, the key evidence against
Bolding was an alleged confession. During a 23-hour interrogation,
Bolding reportedly confessed to killing two patients and agreed to write
letters of apology to their families. She later retracted this
confession, however, and it was excluded from the trial after a judge
found that it had been obtained through illegally coercive methods that
violated Bolding's constitutional rights. Consequently, prosecutors
had little to rely upon during the trial other than the statistical
evidence. No one testified to seeing Bolding inject any patients with
potassium chloride, and although post-mortem examinations showed that
the patients had higher than normal potassium levels, it was impossible
to determine whether potassium chloride poisoning was the cause of the
deaths. Defence lawyers offered alternative theories for the elevated
rate of deaths during periods when Bolding was present.}

\medskip

\emph{A judge, who decided the case without a jury, found the
prosecution's statistical evidence insufficient to warrant a
conviction, saying ``the state at most has placed [Bolding] at
the scene of the offenses\ldots but that is insufficient to sustain a
conviction.'' (Washington Post, June 21, 1988). According to the judge,
the statistical evidence ``failed to supply the missing link that
would connect the defendant with the alleged criminal act,'' and
consequently ``the state's reach hopelessly exceeded its
grasp'' (AP News, June 20, 1988).}
\end{boxedminipage}

\bigskip

Sacks testified at Bolding's trial that ``[the] chances of [this
large number of cardiac arrests] happening by chance is about one in
100 trillion''. Faced with such evidence, it is understandable that
authorities may declare: ``This could not have been a coincidence!'' and
conclude that the only reasonable explanation is criminal misconduct.
Yet a judge acquitted Bolding of charges that she killed three gravely
ill patients and attempted to murder two others by injecting them with
massive doses of potassium chloride, finding this statistical evidence
insufficient to sustain a conviction.

As we will explain, there are a number of potential problems with
computations and interpretations based on such evidence. Relative risk
is difficult to compute in a manner that takes appropriate account of
all relevant variables, and efforts to evaluate relative risk may be
distorted by a variety of biases and predictable errors that can make
such statistics misleading.\footnote{This is a separate issue from that
  of how to report changes in risk. It is important to distinguish
  between relative risk and absolute risk to avoid misinterpretation.
  The absolute risk is the number of events (good or bad) during a
  stated period in treated or control groups, divided by the number of
  people in that group. The relative risk is the ratio of the absolute
  risk in the treatment group, divided by that in the control group.It
  is very common, both in serious studies and in everyday life, for a
  large relative risk to be reported for what may be a very small, even
  negligible, increase in absolute risk. See Appendix 2, and
  Spiegelhalter (2017). See also the BMJ blog article
  \href{https://blogs.bmj.com/medical-ethics/2019/05/14/making-a-pigs-breakfast-of-research-reporting/}{{Making
  a pig's breakfast of research reporting}} which includes the sentence
  "This implies that every single portion of bacon increases risk by
  20\%, when in fact the study found that only increased consumption
  over time increases absolute risk by 0.08\%."}

Recommendations for appropriate statistical analysis are discussed in
Section 5.\footnote{See also the Inns of Court College of Advocacy/Royal
  Statistical Society report on Statistics and Probability for Advocates
  (2017).} We will leave those problems aside for the moment, however,
in order to discuss a more fundamental issue: what conclusions can
reasonably be drawn from a seemingly unlikely cluster of occurrences?

\hypertarget{seemingly-unlikely-coincidences-can-and-do-occur}{%
\subsection{Seemingly unlikely coincidences can and do
occur}\label{seemingly-unlikely-coincidences-can-and-do-occur}}

It is important to acknowledge that seemingly unlikely coincidences
occur regularly -- in other words rare events do happen. The individual
chances of winning a lottery, for example, are often extremely low, yet
winners are declared regularly and it is rare that anyone takes the low
probability of winning as indicating that the victory ``could not have
been a coincidence.'' A California couple once managed to win two
separate lotteries in a single day. According to one estimate, the
probability of winning both lotteries by chance was approximately one in
26 trillion.\footnote{See ABC News,
  \href{https://abcnews.go.com/US/story?id=90981\&page=1}{{California
  couple wins two lotteries in one day}}.} Yet there was no general
belief that this probability was low enough to rule out the possibility
it was merely coincidence; no serious claim that it proved the lotteries
had been ``fixed''. It is therefore worth considering why the low
\emph{p}-value assigned to a relative risk analysis can be taken as
powerful evidence that a medical professional engaged in criminal
misconduct, while a similarly low probability of winning a lottery is
not viewed as convincing evidence that the lottery was corrupt.

The difference may arise, in part, from our understanding that many
people play the lottery. There may only be one chance in many millions
of winning, but if many millions of people play, it becomes quite
plausible that someone will, by chance, be a winner (and this holds even
without the lottery operator engineering wins by eg carrying over
unclaimed prizes to later draws). When suspicions arise against a
medical professional, by contrast, our focus is on that single
individual. We are not thinking about the likelihood that an unusual
number of deaths will be observed somewhere among the patients of one of
the millions of medical professionals in the world; we are thinking
about the likelihood of so many deaths among patients of a particular
individual. While we expect someone to win a lottery; we do not expect
any particular player to win; and by similar logic we do not expect a
rash of deaths among patients of a particular medical professional.

This difference may be more apparent than real, however, particularly if
an apparently anomalous number of deaths is the very reason that
suspicions arose against that individual, or was the reason that such
suspicions led to a criminal prosecution. It is unlikely that a
1-in-10-million coincidence will incriminate any particular medical
professional, but given the very large number of medical professionals
in the world, it is likely, perhaps even inevitable, that such a
coincidence will affect the patients of some medical professional at
some medical facility somewhere in the world.\footnote{For a different
  example making a similar point, it is a simple calculation that if
  1000 events occur completely at random in a year, there is a more than
  93\% chance that one of the 365 calendar days includes 8 or more
  events, yet the chance of 8 or more on one \underline{particular} day
  is less than 1\%.} Consequently, if we take the 1-in-10-million
coincidence to be evidence of medical misconduct, it is inevitable that
we will falsely incriminate innocent people. Thus, the existence of a
cluster of deaths among patients of a medical professional should not,
in itself, be taken as proof of criminality. We are not suggesting that
such evidence is worthless; indeed, as we will explain, it may have
substantial probative value, but its value needs to be assessed
carefully in light of the other evidence in the case.

\hypertarget{the-importance-of-avoiding-illogical-inferences-from-p-values}{%
\subsection{\texorpdfstring{The importance of avoiding illogical
inferences from
\emph{p}-values}{The importance of avoiding illogical inferences from p-values}}\label{the-importance-of-avoiding-illogical-inferences-from-p-values}}

Suppose that an unexpectedly large number of deaths occur among patients
of a particular medical professional. Suppose further that an expert
concludes that the probability of so many deaths occurring by chance is
only 0.000001, or one in a million. What can we conclude about the
chances that the medical professional engaged in misconduct?

As mentioned earlier, there are a number of potential problems
surrounding the computation of statistics of this type, \emph{p}-values,
but let us leave such issues aside for the moment and assume that the
expert's probability is well supported by available data. What can we
conclude from it?

People often think that a \emph{p}-value tells them the probability that
a coincidence occurred. So, it may seem reasonable to assume that the
\emph{p}-value means there is only one chance in a million that so many
deaths occurred by coincidence and correspondingly 999,999 chances in a
million that the deaths arose from some other cause. If misconduct by
the medical professional appears to be the only plausible alternative
explanation, then people might be tempted to conclude that the
probability of misconduct is overwhelming (999,999 chances in 1
million). Thus, the \emph{p}-value of 1 in 1 million might inadvertently
be taken as proof that there is only 1 chance in 1 million that the
medical professional is innocent of misconduct.

Research has shown that people are often tempted to think in this manner
and have difficulty seeing errors in this chain of logic. Yet, this line
of thinking is indeed erroneous and illogical, and may cause people to
misinterpret the value of statistical evidence in ways that are
dangerous and unfair for medical professionals accused of misconduct.

The problem arises from misinterpretation of the \emph{p}-value. The
\emph{p}-value is not the probability that a coincidence occurred.
Rather, the \emph{p}-value is the probability of the observed or more
extreme evidence \emph{if} it arose due to coincidence. In a case
involving unexpected deaths, the expert might ask, for example, ``what
is the probability, by chance, of observing a given number of deaths (or
more) among patients of a particular professional, \emph{if} we assume
that the rate of deaths is no higher than for patients of other similar
professionals?" That is a different question than asking ``what is the
probability that the explanation for these deaths is mere coincidence?''
Importantly, the answer to the former question does not tell us the
answer to the latter, although people may have difficulty seeing why
not.

\bigskip

\begin{boxedminipage}{0.95\textwidth}

\emph{\textbf{Misunderstanding of p-values}}

\medskip

\emph{\textbf{Question expert attempts to answer}:}

\emph{``What is the probability of observing this many deaths (or
more) by chance or coincidence IF the risk of deaths to the
patients in question was in fact no higher than the risk to
similar patients who were not under the care of the accused?''}

\medskip

\emph{\textbf{Question people may mistakenly think expert is
answering}}:

``\emph{What is the probability that the high number of deaths
observed among the patients in question is explained by chance or
coincidence?}''

\end{boxedminipage}

\bigskip

A \emph{p}-value is a statement about a conditional (cumulative)
probability -- that is, a statement about the probability of one event
(observing the data, or something more extreme) in light of another
event (there is no effect of interest, only chance).\footnote{The
  ``another event'' is a hypothesis; in some philosophical traditions,
  this would not be considered an event, but the logic that follows
  holds regardless.} People are generally familiar with the idea that
the occurrence of one event may change our uncertainty about another
event. For example, the probability of rain if the sky is cloudy will
differ from the probability of rain if the sky is clear. But
psychological research has shown that people often fall victim to a
logical fallacy known as ``transposing the conditional'',
misinterpreting the probability of the evidence given a hypothesis with
the probability of the hypothesis given the evidence. In a legal
context, this error has been known as the ``prosecutor's fallacy,"
the ``source probability error,'' or the ``fallacy of the
transposed conditional''.\footnote{Thompson \& Schumann, 1987; Koehler,
  1993; Balding \& Donnelly, 1994; Evett, 1995.} It is a widespread mode
of reasoning, affecting many of the general public, the media, lawyers,
jurors and judges alike.\footnote{For example, in the trial of Sally
  Clark for double infanticide, an expert medical witness testified that
  the probability that both her babies would have died from natural
  causes was one in 73 million (This figure has itself been widely and
  properly criticised, but that is not the issue here). If, as appears
  very natural, we refer to this figure as ``the probability that the
  babies died by innocent means'', it is all too easy to misinterpret
  this as the probability (on the basis of the evidence of the deaths)
  that Sally is innocent -- such a tiny figure seeming to provide
  incontrovertible proof of her guilt. For other evidence of the
  widespread appearance of this fallacy, and the difficulty people have
  in appreciating it, see Thompson \& Newman, 2015; de Keijser \&
  Elffers, 2012.} When incorrectly stating that the \emph{p}-value tells
us the probability a medical professional is innocent of misconduct, the
error of logic is not so easy to see.

The error is to confuse or equate the conditional probability of event
A, given that event B occurs, with the conditional probability of event
B, given that event A occurs; the fault in this logic is easy to
illustrate by inserting various propositions for A and B. Consider, for
example, that the probability an animal has four legs if it is a dog is
not the same as the probability an animal is a dog if it has four legs.
The probability a person speaks Spanish if they grew up in Peru is not
the same as the probability they grew up in Peru if they speak Spanish.
With simple examples like these, the logical error of equating the two
conditional probabilities is easy to see. But this same error underlies
the assumption that the \emph{p}-value tells us the probability a
medical professional is innocent of misconduct, and in that context the
error of logic is not so easy to see. A statistician testifies that the
conditional probability of observing so many deaths (or more) is only
0.000001, or 1 chance in a million, if the deaths are occurring randomly
at rates no higher than among similar patients of other medical
professionals. Here it is tempting to transpose the conditional
probabilities -- to assume that there is also only 1 chance in 1 million
that the deaths are occurring randomly given that so many deaths
occurred.

Of course investigators and triers-of-fact in these cases need to
evaluate whether the deaths in question could have occurred by chance.
While the \emph{p}-value does not answer that question directly, it may
nevertheless be helpful by casting light on the importance of the death
rate when considered in combination with other evidence in the case.

A final important caveat about use of \emph{p}-values is that the
assumptions underlying their calculation include that the test in
question is the \emph{only} test to be conducted. If large numbers of
hypotheses are tested, then some will yield statistically significant
results just by chance -- that chance is precisely what the numerical
\emph{p}-value measures -- so if several tests are conducted some
adjustment for multiple testing should be used.

The correct way to ``invert the conditional'' and avoid the
Prosecutor's Fallacy involves a simple logic that is codified in
what is known as ``Bayes' rule'', that combines the
probabilities of the evidence given various possible explanations for
the data ``hypotheses'') with the prior probabilities of these
hypotheses, to deliver the probabilities of each hypothesis \emph{given}
the evidence. It is the same logic as is used to report a diagnosis
following a medical test. In Appendix 4, Bayes' rule is explained,
using an example presented in non-technical language.

As we will discuss in the following sections, the connection between a
\emph{p}-value and the probability of misconduct by a medical
professional becomes even weaker and more problematic when there are
other possible explanations for the evidence (other than coincidence or
misconduct by a particular individual), and when a \emph{p}-value is
calculated in a biased and misleading manner.

\section{Competing theories}

There are documented cases in which medical professionals have
intentionally engaged in misconduct that put their patients at risk. A
well-known example is that of Harold Fredrick Shipman, an English
physician in general practice (see box below). In 2000, Shipman was
found guilty of the murder of 15 patients under his care.\footnote{See
  The Shipman Inquiry (2003).}

While it is important to acknowledge that cases like that of Shipman
exist, it is also important to realise that they are extremely rare. Of
the hundreds of millions of medical professionals in the world, the
number who are known to have been serial killers of their patients is
very small, a miniscule fraction of the total number. Consequently, in
the absence of any other evidence of misconduct, the prior odds of any
particular medical practitioner engaging in such conduct must be
considered extremely low, of the order of one chance in
millions.\footnote{There is an extensive peer-reviewed literature
  quantifying this risk: see for example Forrest (1995).} As noted in
the previous section, such low prior odds will often be difficult to
overcome on the basis of statistical evidence alone. Even if there is a
cluster of deaths among the practitioner's patients
that is a million times more probable if the practitioner is a murderer
than if the deaths occurred by chance, a logical assessment of the
posterior odds might still conclude that the theory of coincidence is
more probable.\footnote{Posterior odds are defined in Appendix 4.}

\bigskip

\begin{boxedminipage}{0.95\textwidth}

\emph{\textbf{The case of Harold Shipman}}

\medskip

\emph{There are documented cases in which medical professionals have
intentionally engaged in misconduct that put their patients at risk. A
well-known example is that of Harold Fredrick Shipman, an English
physician in general practice. In 2000, Shipman was found guilty of the
murder of 15 patients under his care. Investigators suspected he was
responsible for the deaths of many others, perhaps as many as 250,
making him one of the most prolific serial killers in modern history.}

\medskip

\emph{Concerns about Shipman were first raised by other medical
practitioners, who noted what appeared to be an unusually high rate of
deaths among Shipman's patients. An initial police investigation in
1998 found insufficient evidence to bring charges, but police
subsequently learned that the wills of some of Shipman\'s former
patients had been altered under suspicious circumstances to leave assets
to Shipman, rather than family members of the deceased. Further
investigation found evidence that Shipman had administered lethal doses
of sedatives to healthy patients, and had then altered medical records
to indicate falsely the patients had been in poor health. Based on this
evidence Shipman was prosecuted and convicted.}

\medskip

\emph{In light of this grim episode, there were calls for improved
monitoring of adverse medical outcomes, to allow dangerous medical
misconduct to be detected earlier. For example, statistician David
Spiegelhalter and colleagues suggested that statistical monitoring of
patient death rates would have raised red flags about Shipman's
misconduct years earlier, thereby saving lives (see Spiegelhalter, D.~et
al., 2003).}

\end{boxedminipage}

\bigskip

An assessment of the posterior odds may change dramatically, however,
should other evidence emerge that supports the theory of misconduct,
such as evidence of the altered wills and altered medical records in
Shipman's case. Evidence of this type, when
considered in combination with the statistical evidence, may well make
an overwhelming case in favour of conviction. Consequently, it is vital
that suspicions raised by apparent statistical anomalies be followed by
a careful investigation that looks for other evidence, such as evidence
of motive, consciousness of guilt, or actual lethal medical
interventions. Furthermore, as discussed in later sections, it is vital
that such an investigation be conducted in a neutral, open-minded
fashion to minimise bias. If such an investigation is conducted and no
supporting evidence is found, that may be a strong indicator that the
statistical anomaly indeed arose from coincidence, or that causal
factors other than misconduct by the accused individual are responsible
for what happened.\footnote{Absence of evidence \underline{can}
  sometimes be evidence of absence, see commentary of Aart de Vos,
  translated in Gill (2021), and Thompson \& Scurich, 2018.}

Investigators should always bear in mind that there may be innocent
explanations for apparent (and even striking) correlations between a
particular professional's presence in a hospital on the one hand, and
deaths, resuscitations, or other incidents on the other hand.
Correlations might be caused by many factors, some of which might in
principle be known, but still hard to take account of; some might be
unknown altogether. Seriously ill patients on a medium care ward are
quite likely to die at any moment, but the best medical professionals
will not be able to predict exactly when. In one such hospital
situation, statistical analysis of registered times of deaths shows that
most deaths happen in the morning.\footnote{See Dotto, Gill \& Mortera,
  2022, especially Figs 2 and 3.} The physiological explanation is that
after some sleep, bodily functions resume, and organs close to breaking
point can suddenly fail. In this hospital, there are many nurses on duty
during the morning shift, starting at 7 a.m. and lasting seven hours
till 2 p.m. This is also the period when medical specialists make their
rounds. In the afternoon and evening, there are fewer nurses on duty.
Things are quietening down for the night. There are also fewer deaths
and emergencies, and fewer medical specialists present. During the night
shift, everything is very quiet, there are few nurses on duty. There are
also few events. Contrary to popular imagination, most people do not die
in their sleep at night; they die in the morning while waking
up.\footnote{See Mitler, et al, 1987.}

All this means that most nurses (especially full time, fully qualified
nurses) spend many more hours on duty during the morning, and even less
during the night, compared to the afternoon and evening. Most deaths
occur (or at least, are registered to have occurred) in the morning.
Therefore, most full-time, fully qualified nurses do experience many
more deaths when they are on duty than occur when they are not on duty!

A complicating factor is that time of death has to be registered by a
doctor. The rules on what a doctor is supposed to write on a death
certificate vary in different jurisdictions: is it the time that they
guess that the patient had died, or is it the time at which they sign
off that they have determined that the patient has died?\footnote{Currently
  in England and Wales there are no rules about this; in Scotland it is
  quite clearly codified at Scottish Government, 2018.} In some
countries, data show that ``official'' times of death are often rounded
to whole hours or whole half hours, sometimes even to whole days. A
patient found dead in the morning might be registered as having died at
five past midnight or at five past seven, for all kinds of innocent
reasons.

The activities of nurses can certainly influence these registered times.
One might imagine that a better nurse checks up on all their patients
more often and checks up more carefully and is more aware of how they
are doing. A better nurse will therefore notice and signal a death (or
an emergency) earlier than a worse nurse; thereby causing deaths to be
registered in their shifts and not later. A better nurse will clock-in
for work well before their shift starts, and clock-out well after it
ends, in order to participate fully in the necessary hand-over from one
shift to the next. There is a lot of evidence that the apparent excess
of deaths when Lucia de Berk (see box over page) was on duty was
connected to the fact that she was in fact a better (more hard working,
more conscientious) nurse than many of her colleagues.\footnote{See
  Gill, et al, 2018, Meester, et al., 2007 and Schneps \& Colmez, 2013.}
In fact we know that she had been evaluated as an excellent nurse and
consequently rapidly gained the necessary qualifications to be entrusted
with harder tasks. The rapid increase in deaths on her ward coinciding
with this ``promotion'' also coincided with a management decision to
transfer babies with genetic birth defects from intensive care (where
any deaths occur in different circumstances and are likely to be
accurately recorded) to medium care, in order that they might then be
rapidly transferred to home.

Notice that the nurses themselves, as well as their lawyers, or the
experts they call as witnesses, may not think of these alternative
hypotheses themselves. The same may be true of hospital authorities, who
may be the first to sound the alarm and perform their own
investigations, followed by police investigators and public prosecutors.
Attention may then focus on the possibility that a suspected individual
engaged in nefarious actions; the investigators may lack the knowledge
or, in some instances the motivation, to identify possible alternative
explanations.

There are additional examples of cases in which a cluster of deaths,
initially attributed to individual misconduct, turned out to have
another explanation. A cluster of deaths in a neo-natal ward in Toronto
was initially associated with a nurse, who was suspected of malevolent
activity. Only later was it discovered that new artificial latex
products in feeding tubes and bottles could have been
responsible.\footnote{See Hamilton, 2011.} An apparent increase in death
on a neonatal ward in England raised similar suspicions until a medical
statistician identified the date at which the death rate rose, and a
neonatologist recognized it as the date when the supplier of milk
formula was changed. As these examples show, an increase in deaths may
be caused by factors that are not immediately apparent, even to those
involved. Such factors may require considerable expertise to discover
and could be missed entirely in some instances.

\bigskip

\begin{boxedminipage}{0.95\textwidth}

\emph{\textbf{The case of Lucia de Berk}}

\medskip

\emph{In 2003, Lucia de Berk, a Dutch paediatric nurse, was convicted of
four murders and three attempted murders of children under her care. In
2004, after an appeal, she was convicted of seven murders and three
attempted murders. Thereafter, several academic commentators questioned
the quality of the evidence used to support the conviction, particularly
statistical testimony.}

\medskip

\emph{De Berk had been under suspicion in her hospital for some months
as a result of gossip about her tough, disturbed childhood and striking
personality. When a child in her care died suddenly, the death was
immediately announced to be completely unexpected and, by implication,
suspicious. Hospital officials identified eight further deaths or
resuscitations that had occurred while she had been on duty as medically
suspicious. Additional suspicious deaths were identified and linked to
de Berk at two other hospitals where she had worked. For two of the
patients, investigators found toxicological evidence supporting the
claim that de Berk had poisoned them, although the probative value of
this evidence was weak. Statements in de Berk's diary about
``a very great secret'' and a ``compulsion'' on a day that a
patient had died were given a sinister interpretation.}

\medskip

\emph{During her original trial, a criminologist (who had years earlier
graduated in mathematics) presented statistical evidence according to
which the probability of so many deaths occurring while de Berk was on
duty was only 1 in 342 million. This number was the product of three
p-values, one for each hospital. Prominent statisticians came forward to
argue that the incriminating statistic was based on an over-simplified
and unrealistic model, biased data collection, and a serious
methodological error in combining p-values from independent statistical
tests. The probability of so many deaths occurring by chance may have
been as high as one in 25.}

\medskip

\emph{In light of these doubts, and further medical evidence that came
to light in post-conviction investigations, the case was re-tried in
2010 and de Berk was acquitted. The original convictions are widely
viewed as miscarriages of justice that were prompted, in part, by an
inadequate investigation and misuse of statistical evidence. They led to
various reforms in the Dutch legal system.}

\end{boxedminipage}

\bigskip

\section{Investigative bias}

Because criminal investigations are carried out by human beings,
investigative findings may be influenced by common human tendencies
(often called biases) that can affect the way investigators search for
and evaluate evidence, as well as how they choose to report findings.
This section will discuss ways that various widely recognised human
biases may affect the investigation of misconduct by medical
professionals, with a particular focus on how such biases may affect the
statistical findings. The following section (Section 5) will discuss
ways to minimise such biases to facilitate more objective and useful
investigative outcomes.

\hypertarget{unconscious-bias-throughout-society}{%
\subsection{Unconscious bias throughout
society}\label{unconscious-bias-throughout-society}}

Although a variety of specific biases have been identified,\footnote{See
  Sackett, 1979, and Appendix D to The Law Commission, 2015.} they
generally arise from a common phenomenon: that people's expectations and
desires can influence what they look for and how they evaluate what they
find when they seek answers to important questions.\footnote{Kassin,
  Dror \& Kukucka, 2013; Thompson, 2009; Nickerson, 1998.} The tendency
of preconceptions and motives to influence people's interpretation of
evidence has been called ``one of the most venerable ideas of \ldots{}
traditional epistemology\ldots'' as well as ``\ldots{} one of the better
demonstrated findings of twentieth-century psychology''.\footnote{Nisbett
  \& Ross, 1980.} This tendency, which is often labelled an observer
effect,\footnote{Risinger \emph{et al}, 2003. For the relevance to
  epidemiological methods, see also Sackett, 1979.} was mentioned in
both classic texts and in the writings of early natural philosophers,
such as Francis Bacon, who observed in 1620 that:

\emph{The human understanding, when any proposition has once been laid
down \ldots\ forces everything else to add fresh support and confirmation;
and although \ldots\ instances may exist to the contrary, yet \dots\ 
either does not observe or despises them.}\footnote{Bacon, 1620.}

The potential for observer effects to distort scientific investigations
was recognised by early astronomers, who discovered differences in
reported findings of the same astronomical phenomena by different
observers.\footnote{Risinger \emph{et al}, ibid.} Historians of science
have noted numerous additional ways in which human expectations or
desires may explain incorrect reports of scientific observations, such
as scientists' failure to notice (or at least to report) phenomena
inconsistent with their theory-based expectations; reported findings
that support pet theories but cannot be replicated; and the
statistically improbable degree of correspondence that has been observed
between some reported findings and theoretical expectations (the most
famous example from the history of science being the improbable degree
of agreement between data and theory in Gregor Mendel's experiments with
peas).\footnote{Pires \& Branco, 2010; Jeng, 2006.} Over the last 70
years, epidemiologists and statisticians have developed methods to limit
and assess the impact of many of these biases.\footnote{Hill, 1965, is
  an early seminal reference. For approaches useful in forensic science,
  see Stoel et al., 2015, Dror et al., 2015.}

The same scope for biased data collection has been noted in criminal
investigations. Miscarriages of justice are often attributed to ``tunnel
vision'' and ``confirmation bias,'' processes that may lead
investigators to ``focus on a particular conclusion and then filter all
evidence in a case through the lens provided by that
conclusion.''\footnote{Findley and Scott, 2006, p. 292.} The comments of
Findley and Scott on the underlying process are reminiscent of what
Francis Bacon (quoted above) wrote in 1620:

\emph{Through that filter, all information supporting the adopted
conclusion is elevated in significance, viewed as consistent with the
other evidence and deemed relevant and probative. Evidence inconsistent
with the chosen theory is easily overlooked, or dismissed as irrelevant,
incredible, or unreliable.}\footnote{Findley and Scott, ibid.}

Common investigative practices noted in the commentary by Findlay and
Scott include tendencies for investigators:

\begin{itemize}
\item
to settle too quickly on a preferred theory, without adequately
considering alternatives;
\item
to look for evidence that confirms or supports the preferred theory
rather than seeking evidence that might disconfirm it or support
alternatives;
\item
to notice, remember and record evidence more readily and reliably when
the evidence is consistent with the preferred theory than when it is
not;
\item
to interpret ambiguous evidence in a manner consistent with the
preferred theory;
\item
to view evidence and interpretations as more credible when they support
the preferred theory, and \emph{vice versa};
\item
to report findings with a higher degree of confidence if they support
rather than contradict the expected result;
\item
to fail to hand over or disclose all the countervailing evidence to the
defence; and
\item
  to have skewed incentives to boost their case.
\end{itemize}

A key part of any investigation is attempting to identify the full range
of hypotheses that need to be explored, though it is difficult to be
confident this is done successfully. Investigations can go awry, and
produce misleading findings, if investigators miss or ignore an
underlying causal factor. Deadly misconduct by a medical professional is
one possible explanation for an unexpected surge in the death rate at a
medical facility. Coincidence is another possible explanation (as
discussed in Section 2(a)): clusters do occur by chance even in a
completely random pattern of events. If investigators assume misconduct
and coincidence are the \underline{only} plausible explanations, then
evidence indicating that coincidence is unlikely will lead investigators
inevitably to the conclusion that misconduct is likely. This conclusion
may be mistaken, however, if the elevated death rate arose, even in
part, from other confounding factors, such as changes in the underlying
population of patients; negligence or misconduct by other individuals;
administrative changes affecting such matters as staffing levels,
training, or case load; or medical policy changes affecting transfer of
patients from one hospital section to another. See also the discussion
on competing theories in Section 3.

The difficulty of identifying possible causal factors is
multi-faceted.\footnote{The methods developed in medicine to bring
  together evidence from laboratory science, observational and
  experimental studies are important tools for investigation. In the UK,
  the Forensic Science Regulator has published advice on the development
  of evaluative reporting.} It may arise in part from basic human
psychological tendencies, as well as from self-interest of individuals
involved. Psychological research suggests that people have a general
tendency to gravitate toward criminality as an explanation for seemingly
anomalous events, rather than looking at situational or institutional
factors; this is called the ``fundamental attribution
error''.\footnote{Nisbett \& Ross, 1980; Ross, 1977.} It causes people
to look to the person (ie, to human agency) rather than the situation
when explaining events. There is strong and well-demonstrated
psychological tendency for people to assume that bad things are caused
by bad people rather than bad circumstances (cf. the common public need
to attribute blame to individuals, for ``heads to roll'', in cases of
systemic failure in public services).\footnote{Burger, 1981; Ross, 1977.}
Hence, people may tend to look for scapegoats to blame for bad medical
outcomes arising from other causes, and this is often encouraged by
sensationalist reporting in the media.

The tendency toward scapegoating may be abetted by stereotyping and
bias. Individuals accused of medical misconduct have often been unusual
in ways that drew attention and ultimately suspicion, making the
hypothesis of medical murder appear more plausible than otherwise. While
it is important for investigators to take account of suspicious
behaviour when that behaviour is diagnostic of misconduct, a focus on
the odd-ball or iconoclast may unfairly distort investigators'
impressions of the matter if they mistakenly rely on stereotypes that
have little probative value. Stereotypes with no empirical support, such
as the notion that nurses whose fashion aesthetic tends toward the
``gothic'' are more likely than other nurses to commit murder, may draw
unwarranted suspicion to certain individuals and make them investigative
targets. People are not always conscious of the effect of such
stereotypes on their thinking,\footnote{Nisbett \& Wilson, 1977.} and
through this unconscious bias, statistical evidence offered against a
Goth nurse may be taken more seriously and examined less critically,
than the same evidence would be if it were offered against a more
conventional individual.

Cognitive biases can also affect the way that investigators interpret
and classify data, and thereby distort the findings that emerge from an
investigation. Epidemiological and statistical methods used in
investigations of disease outbreaks or clusters of adverse events are
applicable to investigating clusters of deaths.

Whether a particular death should be deemed ``suspicious'', for example,
might be influenced by a variety of factors, including factors that have
little or no diagnostic value. Cognitive psychologists have found that
people often have limited insight into the factors that influence such
evaluations, so can be influenced by their own expectations or motives
without realising it.\footnote{Nisbett \& Wilson, ibid.} The largely
unconscious nature of these processes makes the resulting biases
difficult to remedy. Teaching people about such biases is not sufficient
to prevent them, nor is exhorting people to be unbiased.\footnote{Kassin
  \emph{et al}., 2013; Risinger \emph{et al}. 2003.} The most reliable
and effective counter-measure is actively to avoid creating the biases
in the first place by arranging, to the extent possible, to avoid
creating strong expectations of desires for a particular
outcome.\footnote{Dror \emph{et al}., 2015.} Additionally one needs
proper checks and balances including ``red-teaming'' and independent
stringent review of potential evidence, as done, for example, by the CPS
in England and Wales.

\hypertarget{anatomy-of-a-biased-investigation}{%
\subsection{Anatomy of a biased
investigation}\label{anatomy-of-a-biased-investigation}}

The general points about investigative bias offered in paragraph (a)
above allow us now to consider more specifically ways that bias may
distort the investigation of medical professionals accused of harming
patients. We will describe a hypothetical (but not completely fanciful)
case in which a medical professional is accused of mass murder; we will
then discuss how the cognitive tendencies discussed above may lead that
investigation awry.

Let us suppose that administrators at a hospital become aware of an
alarming increase in the number of deaths among elderly patients.
Suspicion falls on a doctor who works in a unit where many deaths
occurred. The doctor has drawn attention by speaking publicly in favour
of euthanasia, making comments suggestive of relief rather than sadness
after some patients died, and by appearing at the hospital's Halloween
costume party dressed as the Angel of Death. When a co-worker reports
seeing a syringe in the doctor's bag, hospital administrators call for
an investigation.

At this stage, investigators often seek to determine whether the surge
in deaths can be linked to the suspect. Were patients more likely to die
when this doctor was on duty than when other doctors were on duty? To
address this question, investigators often try to count the number of
deaths for which the suspect may bear responsibility and compare it to
the number of deaths that occurred under similar circumstances when the
doctor could not have been involved.

In order to perform such an analysis, investigators must make a number
of difficult judgments. They must determine, for example, whether each
death that occurred should be viewed as a possible homicide, and if so
whether the doctor in question could have been responsible for that
homicide. Crucially, they must also consider whether factors other than
the presence or absence of the doctor in question may have affected the
rate of death.

Because there is a high degree of subjectivity in such judgments, they
may be subject to bias. Furthermore, as we will explain, if the
investigators are focused on a particular suspect, it is likely that
these biases will slant the investigators' findings in a direction that
is unfairly incriminating to that suspect. The remainder of this section
will discuss ways bias may arise in an investigation; the following
section will discuss possible ways to mitigate such biases.

\hypertarget{suspicious-deaths}{%
\subsection{\texorpdfstring{``Suspicious
deaths''}{``Suspicious deaths''}}\label{suspicious-deaths}}

One important judgment is about which deaths to count. Investigators
typically try to rule out deaths that are readily attributable to known
causes and instead focus on deaths that are ``suspicious'' or
``unexpected'' -- ie, deaths that might possibly have been the result of
homicide rather than disease or other ``natural causes.'' Distinguishing
the former from the latter is a matter that requires expert judgment by
specialists, such as forensic pathologists and researchers who study
death certificate coding.\footnote{Ideally, forensic pathologists would
  be able to examine each corpse, but more realistically, death

  certificates should at least be reviewed by two people independently.}
Research indicates, however, that forensic pathologists sometimes
disagree about manner of death (eg, whether by homicide or accident) and
that such judgments can be influenced by non-medical contextual
information.\footnote{For an example of the need to understand coding of
  deaths see the article on violent child deaths by Sidebotham et al,
  2012.} . Post mortem evidence also confirms a high error rate in
reported cause of death, even for good doctors.

For example, Dror et al recently reported that forensic pathologists who
were asked to evaluate autopsy findings in a hypothetical case were more
likely to conclude that a child died due to homicide rather than
accident if they were told it was a black child under the care of the
mother's boyfriend than if told it was a white child under care of the
child's grandmother.\footnote{Dror \emph{et al}., 2021a.} Whether this
finding reflects ``bias'' is controversial.\footnote{Arguably, the
  experts were asked the wrong question: if they had been asked not
  about cause of death, but about likelihood ratios then context could
  be appropriately separated.} Some commentators have argued that
information about the child and its caretaker is relevant to forensic
pathologists' determination of manner of death and hence that it was
perfectly proper for them to be influenced by it.\footnote{Peterson
  \emph{et al}., 2021.} Dror and his colleagues have argued in response
that the information is beyond what forensic pathologists should
consider when making such determinations and deserves little to no
weight even if it is considered, and hence that their findings are
indeed examples of ``contextual bias''.\footnote{Dror \emph{et al}.,
  2021b.}

For present purposes, the key finding of the Dror et al. study is that
forensic pathologists' manner-of-death determinations can be influenced
by contextual information, such as information about who was caring for
the decedent. Let us consider how that might affect the fairness of the
kinds of investigations we are discussing here. Suppose, for example,
that a forensic pathologist is more likely to determine that a patient's
death was ``suspicious'' and hence possibly due to homicide if aware
that the patient was under the care of a suspected serial killer. This
might happen because the forensic pathologist thinks it is proper and
appropriate to consider such information when evaluating cause of death.
Even if the forensic pathologist tries to ignore such information,
however, it may still bias the evaluation by creating an expectation of
homicide when the pathologist reviews cases associated with the
suspected serial killer, and it may do so without the pathologist being
aware of it.

Contextual information of this kind may also affect thresholds for
reporting. Concern about missing possible victims may cause them to
lower their threshold for reporting ``possible homicide'' when
evaluating patients attended by the suspect; while concern about casting
suspicion on an innocent person causes them to raise the reporting
threshold for patients attended by other nurses. Consequently, when
their evaluation of the medical evidence leaves them uncertain, forensic
pathologists may be more likely to report a case as a possible homicide
if they know the nurse on duty was a suspected serial killer, and less
likely if another nurse was on duty.

Regardless of how it occurs, this kind of bias would undermine the
fairness of the investigation by causing an increase in the count of
``suspicious'' deaths associated with the nurse. The higher count would
arise from the very suspicions that the investigation is supposed to
evaluate -- an example of circular reasoning. Potential remedies for
such biases will be discussed in Section 5.

\hypertarget{access-and-opportunity}{%
\subsection{Access and opportunity}\label{access-and-opportunity}}

Another important judgment that investigators must make is which
suspicious deaths to count ``against the suspect'' (ie, as possible
homicides committed by the suspect) and which to attribute to other
causes. In order to make this determination, investigators need to
evaluate, for each ``suspicious death,'' whether the suspect had, or may
have had, sufficient access to be responsible. That evaluation requires
consideration of a number of factors, such as how the death may have
been caused, how long it would have taken to perform acts that would
cause the death, who else might have been present and whether they would
have observed and reported such misconduct, how soon thereafter the
death would have occurred, how soon it would have been detected, and so
on.

Like judgments about manner of death, judgments about access and
opportunity to kill are complex subjective assessments on which
different experts may have differing opinions (and, where experts are
party-appointed, have implicit ``advocacy-bias'').\footnote{Murrie et
  al., 2013.} Hence, they are also the kinds of judgments that may be
influenced by contextual bias. There is a risk that investigators will
be influenced by their expectations and desires. It is possible, for
example, that they will cast a wider net when looking for ``suspicious
deaths'' that can be linked to a suspect; and a narrower net when
counting suspicious deaths that occurred when the suspect was not
present. As a consequence, the deaths counted against the suspected
individual could increase (relative to deaths counted against others)
for the very reason that the suspect has come under suspicion.

\hypertarget{similar-circumstances}{%
\subsection{Similar circumstances}\label{similar-circumstances}}

To determine whether an unusual number of deaths occurred when a
particular healthcare worker was on duty, it is necessary to compare the
death rate when the worker was on duty with the death rate during
\emph{\textbf{otherwise comparable periods}} when the worker was not. It
is often difficult to make such comparisons in a fair manner, however,
because the presence or absence of the worker may be correlated with
other factors that also affect the rate of death. In other words, the
periods chosen for comparison may not afford a fair comparison with the
periods when the worker was on duty.

Suppose, for example, that the worker typically works the morning shift.
Investigators could compare the rate of deaths that occurred on mornings
when the worker worked with the rate of deaths during the afternoon or
night shifts on the same ward, but the result would be misleading if the
rate of deaths is generally higher during the morning shift than during
the other shifts. Investigators could instead compare the death rates on
those mornings when the worker did and did not work, but that comparison
may also be confounded by other factors. If the worker tended to work
weekdays, for example, and not work weekends, it would be important to
consider whether that factor (weekday vs. weekend) might also make a
difference. Past investigations have sometimes compared rates of death
during the period when a particular individual was on a medical staff
with rates before or after. This kind of comparison confounds that
presence and absence of the individual in question with the period of
time, which could be misleading if time-related changes in procedure,
staffing levels, patient population and the like might also have
influenced death rates. Media reporting of poor outcomes in a hospital
may deter future patients from seeking treatment there, changing the
case mix and influencing future outcomes, and further confounding such
comparisons.

Other factors affecting simplistic comparisons are the possibility of
seasonal effects on disease prevalence and severity, and purely
administrative matters such as the effects of hospital practice on
recording of times of death, presence of doctors, shift changes, etc. In
some hospitals, deaths occurring during a night shift are officially
recorded only in the presence of a doctor at the beginning of the
morning shift.

Investigators must carefully consider such factors in order to make a
fair comparison. That will typically require considerable knowledge
about factors that influence death rates, such as all those identified
above. That suggests that investigators will either need to be
knowledgeable medical professionals themselves or will need guidance
from such professionals.

However, this guidance may itself introduce further biases, when it is
obtained from administrators and staff of the institution being
investigated. Then the officials who guide the investigation may have an
interest in supporting particular outcomes, which could hinder the
ability of investigators to identify the full range of possible causal
factors. Suppose for example, that the increase in deaths that prompted
the investigation arose after administrative changes that affected staff
levels, training, or supervision. To avoid any implications of
responsibility for a surge in deaths, the administrators may well prefer
that the investigation focus on a single bad apple on the staff, rather
than these background factors, and hence may de-emphasise or ignore
them. This self-interested guidance may prevent investigators from
recognising causal factors that confound their assessment of the rate of
deaths attributable to a particular individual. It would be far better
if the investigators had access to sophisticated guidance from experts
on medical issues and hospital procedure who are independent of the
staff and administration of the institution being investigated.

As in science, we should only compare like with like. When we cannot
guarantee this by the design of the study, it is important to
\emph{control} for differences in other plausible causal factors.

In Appendix 5 we give two hypothetical examples that illustrate ways in
which investigative bias may distort statistical evidence emerging from
investigations of medical misconduct, emphasising that even very small
biases can completely transform the strength of the evidence from weak
to compelling. The second example also demonstrates that failure to
control for differences in a causal factor can also lead to huge biases.

\hypertarget{the-role-of-chance}{%
\subsection{\texorpdfstring{The role of chance
}{The role of chance }}\label{the-role-of-chance}}

Throughout this report, we have recognised that the number of deaths
observed while a particular medical professional is on duty is
influenced by chance, in the form of natural sampling variation:
comparing one period to another, the numbers of deaths will differ
purely due to coincidence. Coincidental fluctuations from population
means are more likely with smaller samples, where the law of large
numbers does not dominate, than larger samples, hence there is a greater
chance of observing an unrealistically high or low number of deaths for
shorter intervals than for longer intervals, and for smaller patient
populations than for larger ones. Calculations of statistical
significance are precisely answering the question of whether observed
differences between periods are greater than can reasonably be
accommodated by these chance effects.

However, there are many other reasons why even in the absence of a
causal effect of a medical professional's actions, numbers of deaths in
different periods will differ. The behavioural aspects of these reasons
have been grouped together by Kahneman into what he calls
``noise''.\footnote{Kahneman, Sibony and Sunstein, 2021.} This is really
an umbrella term for quite different kinds of effect that we prefer to
distinguish, as they require different treatments.

We have already discussed that other measurable factors such as season,
time of day, etc., that differ between periods must be included in the
analysis before effects can be attributed to particular causes, and we
have illustrated how to do this. However, in addition, we have to accept
the possibility of \emph{unmeasured} confounders, the ``unknown
unknowns'' we mentioned earlier. Statisticians often accommodate these
factors probabilistically, for example, using random effect models,
bringing a second role of chance into the situation, one for which the
law of large numbers does not assist us (unless we are dealing with a
large number of periods). A simpler approach is to allow for
overdispersion in standard models, assuming for example ``extra-Poisson
variation''.

While the \emph{p}-values presented in these examples take into account
variability due to the size of the samples (ie, the number of cases
examined), they do not, and cannot, take into account additional
variability that may arise due to unmeasured variables. As discussed
earlier, such factors may well increase or decrease the rate of deaths
observed in actual cases. That means that bad luck (due to
investigators' failure to appreciate such factors) may increase the
number of ``suspicious deaths'' that are counted against an innocent
suspect in such an investigation, quite independently of the factors
discussed in the hypothetical illustrations; and good luck (if that is
the proper term) may decrease the number of deaths that are counted
against a guilty suspect (if, for example, a killer happened to commit
murders during periods when an unappreciated variable caused the death
rate to be low).

\section{Advice on investigative procedures}

The foregoing analysis leads to three pieces of advice for professionals
who are asked to investigate alleged misconduct by medical
professionals. First, do not be oblivious to other factors that may
affect the negative outcomes under investigation. Try to understand
fully the factors that may have affected the rates of death or other
negative outcomes that are at issue, and try to take all of those
factors into account when assessing the likelihood that negative
outcomes arose from misconduct by a particular individual.

Second, do not be biased, indeed take active steps not to be. Be
familiar with the potential for cognitive bias and the subtle and often
unconscious ways it can influence expert judgments, and take steps to
minimise such biases. As explained below, that will typically require
the lead investigators to establish context management
procedures,\footnote{See Dror et al., 2015; Stoel et al., 2015.} in
order to control the flow of investigative information to other members
of the investigative team. Failure to take such steps may permanently
impair the investigators' efforts to produce findings that will be
helpful, rather than misleading, to police, prosecutors and
triers-of-fact.

Third, be cautious about drawing conclusions from limited samples, such
as death rates over short periods of time. When examining rates of
unexpected deaths (or other negative outcomes), seek the advice of
statisticians on the appropriateness of the samples selected for
evaluation, on ways to reduce sampling error and various forms of bias,
and on the meaning and proper interpretation of statistical findings (in
terms of both statistical significance and effect size, eg, relative
risk).\footnote{Refer to Appendix 2.}

\hypertarget{identifying-all-potential-causal-factors}{%
\subsection{Identifying all potential causal
factors}\label{identifying-all-potential-causal-factors}}

A key part of any investigation is identifying the full range of
hypotheses that need to be explored. As discussed in the previous
section, investigations can go awry and produce misleading findings, if
investigators miss or ignore an underlying causal factor. If an
investigation is premised on the assumption that an unexpected set of
patient deaths either resulted from intentional misconduct by a given
individual or from coincidence, for example, then evidence that
coincidence is unlikely will appear strongly incriminating. This
conclusion may be misleading, however, if the elevated death rate arose,
even in part, from other factors, such as:
\begin{itemize}
\item
changes in the underlying population of patients,
\item
negligence or misconduct by other individuals,
\item
administrative changes affecting such matters as staffing levels,
training, or case load, or
\item
changes in policy on moving patients between sections of the hospital.
\end{itemize}

Police are often called into such investigations by medical authorities
who become suspicious of a given individual. The police may in turn rely
upon those authorities to familiarise them with the situation and help
them identify possible hypotheses in need of investigation. A danger
inherent in this process is that medical authorities may have an
interest in the outcome of the investigation that influences what they
tell the police about possible causal factors. For example, faced with
an upsurge in patient deaths, hospital administrators may find it easier
to imagine that it was caused by individual misconduct of a ``bad
apple'' on the staff than to acknowledge that it may have arisen from
administrative decisions related to staffing and service levels; this
may happen unconsciously. Identifying possible causal factors is a
complex task requiring considerable expertise both in medicine and
medical administration; it should not be left to amateurs, but also not
left to parties involved in the matter.

Consequently, it is essential that investigators in such cases have
guidance from experts who are independent of the institution being
investigated. Because several types of expertise are relevant, it will
often be necessary to have an advisory team or panel. To investigate a
surge in deaths in the geriatric wing of a hospital, for example,
investigators may need to consult experts in geriatric medicine, experts
in forensic pathology and forensic toxicology, and experts familiar with
hospital procedures, staffing practices and similar issues. The advisory
panel should have sufficient expertise to identify every factor that
might plausibly have contributed to the surge in deaths.\footnote{Notice
  there may even remain ``unknown unknowns''. For example, with hospital
  baby deaths, nobody could have imagined that a change from rubber to
  plastic could have put digoxin-related substances into babies' bodies
  (see Hamilton, 2011). Of course, you can hardly take account of
  unknown unknowns quantitatively. But you should be aware that they are
  possible, and some statistical methods allow approximate adjustment
  for ``unmeasured covariates''.} The advisory panel should also have
the expertise needed to guide investigators on where to look for
evidence that might support or undermine all plausible causal theories.

If the investigators attempt a statistical analysis, it is essential
that they have guidance from a competent statistician. We discuss this
further in paragraph (c) below.

\hypertarget{minimising-bias}{%
\subsection{Minimising bias}\label{minimising-bias}}

In 2009, the United States' National Academy of Sciences observed that
``forensic science experts are vulnerable to cognitive and contextual
bias'' that ``renders experts vulnerable to making erroneous
identifications.'' (p.4 note 8). In the United Kingdom, the Forensic
Science Regulator reached similar conclusions.

Support for these conclusions can be found in a growing body of research
showing that forensic examiners can be influenced by contextual factors
that are irrelevant to the scientific assessments they are supposed to
be performing. For example, latent print examiners who were told that a
suspect had a solid alibi were less likely to conclude that a latent
print found at the crime scene had come from the suspect.\footnote{Dror,
  2006; Dror \& Charton, 2006; Dror \& Rosenthal, 2008.} Similar
findings have been reported in other forensic disciplines, including
document examination,\footnote{Miller, 1984; Stoel, Dror \& Miller,
  2014.} bite mark analysis,\footnote{Osborne, Woods, Kieser \& Zajac,
  2014.} bloodstain pattern analysis,\footnote{Taylor, Laber, Kish,
  Owens \& Osborne, 2016.} forensic anthropology\footnote{Nakhaeizadeh,
  Dror \& Morgan, 2013.} and DNA analysis.\footnote{Dror \& Hampikian,
  2011; Thompson, 2009.}

To minimise such biases, academic commentators and some advisory bodies
have recommended that bench-level forensic scientists adopt context
management procedures that shield them from exposure to potentially
biasing contextual information that is not needed for their scientific
analyses.\footnote{Dror et al., 2015; Thompson, 2015; Thompson, 2011;
  Risinger et al., 2002.} Generally, this requires dividing duties
between a case manager and forensic examiners. The case manager
communicates with criminal investigators, determines what evidence needs
to be collected and what examinations are necessary, and then passes the
evidence on to forensic examiners, who evaluate the evidence and draw
conclusions. The division of duties makes it possible for the case
manager to be fully informed about underlying case and familiar with the
context, while the examiner receives only the information needed to
perform the analysis requested. In this manner the examiners can be
``blind'' to potentially biasing contextual information until after they
have drawn conclusions. The fingerprint examiner does not learn, for
example, whether the suspect has a solid alibi (or not) until after
comparing the prints and drawing a conclusion. This procedure assures
that the examiner's conclusion is based solely on the analysis of the
physical evidence submitted for examination and is not biased by other
contextual information.

We recommend that a similar procedure be employed when investigating
allegations of misconduct by medical professionals. A lead investigator
could play a role similar to that of the case manager by communicating
with other investigators and relevant authorities and identifying
evidence in need of further examination. The lead investigator would be
fully informed of the facts surrounding the case. The lead investigator
would be supported by other individuals with specialised expertise who
would conduct ancillary investigations. These ancillary investigators
would not be fully informed about the underlying case but would
deliberately be kept blind to information that is unnecessary to a fair
scientific assessment, including the prior opinions and conclusions of
other parties in the case.

Consider cases where a medical professional is suspected of killing
patients by poisoning. It will likely be necessary to have toxicologists
examine medical specimens to assess whether deaths that occurred during
relevant periods could have been caused by poison. The toxicologists
will need access to the specimens and will likely need information about
the circumstances under which they were collected, but they need not
know whether the specimens were collected from patients to whom the
alleged killer had access, or from patients to whom the alleged killer
did not have access. Information about access, although potentially
biasing for reasons discussed in Section 4, is clearly irrelevant to the
scientific analysis of the specimen. By keeping the toxicologists blind
to this information, the lead investigator can greatly reduce the risk
that contextual bias will distort the toxicological findings.

Lead investigators will typically need the assistance of forensic
pathologists when assessing whether particular death should be regarded
as ``suspicious,'' and hence as a potential homicide. As with
toxicologists, however, the forensic pathologists need not be fully
informed of all details of the investigation. When assessing whether a
particular death was ``suspicious'' they should remain blind to whether
the suspected medical professional had access to the deceased at least
until after they have recorded their conclusions on cause and manner of
death. Blinding will prevent the kinds of biases discussed in Section 4.

It is possible that some forensic pathologists will resist the use of
blinding procedures that deprive them, even temporarily, of contextual
information. Forensic pathologists in the United States have
vociferously opposed the suggestion that they employ context management
procedures in routine practice on grounds that the case context is
always potentially relevant to their assessment of the medical history
of the deceased and that no one other than a forensic pathologist has
the competence to assess which parts of the case context may be
medically relevant.\footnote{See Simon, 2019.} Additionally, forensic
pathologists are accustomed to receiving all case information when they
assess cause and manner of death for the purpose of issuing death
certificates. For that purpose, they are expected to consider all the
evidence surrounding the case that may bear on how the death occurred,
including circumstantial evidence. Forensic pathologists might even
consider it perfectly proper to take account of the fact that the
deceased was under the care of a suspected serial killer when deciding
whether to report the death as a homicide on the death certificate.

For reasons discussed in Section 4, however, in giving evidence to a
court of law on the specific evidence in question in this kind of
investigation, it is clearly improper for a forensic pathologist to be
influenced by such information. As illustrated in the hypothetical
examples, forensic pathologists may create an unfair bias against
suspects if they allow such information to influence their judgments,
and this undermines the fairness of the legal process. It is for the
court to consider contextual information, such as that the deceased was
under the care of a suspect, and the expert witness's assessment should
be limited to the specific scientific evidence concerned; if this
principle is not adhered to, there is a risk of ``double-counting'' of
the contextual information when the court weighs all of the evidence,
with prejudicial consequences. Biases of this kind are difficult to
control because they can occur without conscious awareness. Simply
instructing a forensic pathologist to ignore contextual information may
well be insufficient; better practice would be to avoid exposing them to
potentially biasing information until after they have made their
assessment.

In summary, blinding is central to proper and defensible conclusions
from analysis of numerical evidence. To promote its increased and
widespread use in legal practice we recommend that it is adopted
wherever practicable, that to encourage good practice its adoption is
disclosed and indeed emphasised in court, and that if it is not
practicable that reasons for this are explained.\footnote{There is some
  evidence that the use of blinding procedures enhances the credibility
  of forensic evidence, particularly when the trier-of-fact appreciates
  that the evidence may depend, in part, on an expert's subjective
  judgments; Thompson \& Scurich, 2019.}

In such investigations a little statistical knowledge may be a dangerous
thing. Medical professionals or social scientists who have taken a few
statistics courses may know how to compute statistics such as relative
risk ratios and \emph{p}-values, but may lack the sophistication and
experience needed to do so in a manner that takes into account all
relevant variables. The statisticians who advise such investigations
should have three qualifications: (1) doctoral level training in
statistics, ideally with an appropriate professional qualification (eg
CStat in the UK); (2) experience with statistical analysis of medical
data; and (3) familiarity with the academic literature on the
investigation of misconduct by medical professionals, such as the
literature cited herein, and this report itself. It is particularly
important that statisticians advising such investigations be familiar
with critical commentary that has identified flaws and limitations in
previous investigations; the reference list contains a few examples. To
improve the quality of future investigations, it is vital that
investigators, and their advisors, treat past mistakes as opportunities
for learning rather than defensively or for blame.''\footnote{See
  Matthew Syed, 2015.}

Some annotated examples of correct analyses of illustrative data on
patterns of occurrence of adverse events are explained at length in
Appendix 6, with explicit computer code to reproduce these in Appendix
8. These analyses make use of the standard statistical methodology of
\emph{log-linear} \emph{models} (taught in the UK in many undergraduate
courses)\emph{.}

\hypertarget{the-role-for-statistics-in-other-specialist-evidence}{%
\subsection{The role for statistics in other specialist
evidence}\label{the-role-for-statistics-in-other-specialist-evidence}}

In this report we concentrate on cases where the data underlying the
evidence in question consists of numbers of events, typically deaths.
However, there is often a need for statistical interpretation of other
kinds of data, typically presented by specialists of other disciplines,
for example sciences such as toxicology or pathology, or social sciences
such as criminology and forensic psychology.

In an investigation where conclusions drawn by specialists also involve
data subject to any variation or uncertainty, a statistical analysis is
necessary; it is important to have the opinion of a statistician about
the soundness of that analysis. Scientific ``intuition'', rules of
thumb, ``standard lab practice'', etc., are no substitute for an
analysis that is correct. As the example in footnote 10 illustrates,
intuition can be a poor guide in evaluating probabilities, especially
regarding apparent ``patterns'' in irregular data.

For an example of a scientific analysis in a criminal case, it is common
in cases where there is an unexpected number of deaths in a hospital
ward that a pathologist is called to assess the cause of death. In the
case of nurse Bolding, and others, the prosecution declared that
potassium chloride was used by the suspect to kill a patient, on the
basis of expert forensic pathologist evidence using earlier scientific
studies that related potassium ion (K\textsuperscript{+}) concentration
in the vitreous fluid of the eye to the post mortem interval (PMI). In a
recent case the relationship between K\textsuperscript{+} concentration
and PMI was used to predict what the ``standard'' amount of
K\textsuperscript{+} should be at a certain (observed) PMI.\footnote{See
  Dotto, Gill and Mortera, 2022.} The patient had a higher concentration
than predicted by the pathologist and the nurse was then accused of
having injected the patient with potassium chloride, causing her death.
In a first trial, this evidence was sufficient to convince the court to
convict the nurse and sentence her to life imprisonment, but the
analysis used to predict what the post-mortem potassium level should
have been was flawed. Not only was the sample used to make the
prediction not representative of the case under examination, but no
measure of uncertainty was attached to the prediction. If a correct
statistical analysis had been carried out the ``incriminating" result
would have been seen to lie within the ``margin of error" -- which means
that the post-mortem analysis did not support the hypothesis that the
dead patients had been injected with potassium chloride. The nurse was
acquitted in a second trial in which a more complete statistical
analysis was presented.

Likewise, any statistical analysis of data generated using other
scientific disciplines should be accompanied by expert opinion in those
disciplines.

\section{Advice for evidence evaluation and case presentation}

\hypertarget{the-lawyers-role}{%
\subsection{\texorpdfstring{The lawyer's
role}{The lawyer's role}}\label{the-lawyers-role}}

Put starkly, a lawyer's role in an adversarial system is to advise and
represent those who instruct them, to the best of their ability. Lawyers
work (or ought to work) from facts to a solution. The advisory role
requires an objective, independent evaluation of the evidence. Clients
are ill-served if they are told only what they want to hear; others
(including those accused of crimes) and the administration of justice
itself may also be harmed by biased reasoning.\footnote{In \emph{The
  Scout Mindset}, Julia Galef describes biased, or directionally
  motivated reasoning as the ``soldier mindset''. By contrast, the
  reasoning underlying the ``scout mindset'' is motivated by accuracy.}
The representative role, however, is in essence the attempt to persuade
the decision-maker (whether judge or jury) to make a finding favourable
to the client on the evidence and argument, and within professional and
ethical boundaries. The dividing line between persuasive and biased
reasoning may be a fine one, and not always easy to identify; the
potential harms of biased reasoning in a representative capacity are,
however, self-evident and ought to be eliminated insofar as possible.

Both advisory and representative roles require an intimate knowledge of
the facts and, where statistics are involved, the uses to be made and
limits of such evidence. We suggest that the starting point ought to be
a full, complete chronology (or thematic organization) of the facts;
this is common practiced in some area of law. Once that has been done,
elementary strategies such as asking ``what do we know?'', ``what don't
we know?'', and ``what do we need to know?'' ought to help to identify
gaps, guide investigations, frame issues, and assist the building of
argument.\footnote{Amy E Herman\emph{, Visual Intelligence}, pp153-164.}
If, once investigations are complete, gaps remain, then care ought to be
taken not to fill them with speculation or (which may be the same thing)
over-determined statistical analysis.

Even once the analysis is complete, there is still a place for
sense-checking. Critical thinking techniques such as the ``double
standard test'', the ``outsider test'', the ``conformity test'', and the
``selective skeptic test'' are likely to be useful in testing inference
and argument.\footnote{Julia Galef in \emph{The Scout Mindset}, p71,
  describes these tests as follows:

  The Double Standards test is ``Are you judging one person (or group)
  by a different standard than you would use for another person (or
  group)''

  The Outsider test is ``How would you evaluate this situation if it
  wasn't \emph{your} situation?''

  The Conformity test is, ``If other people no longer held this
  situation, would you still hold it?''

  The selective Sceptic test is ``If this evidence supported the other
  side, how credible would you judge it to be?''

  Other strategies are to be found in Tom Chatfield's \emph{Critical
  Thinking}.}

Courts in the UK have warned, at the highest level, that ``there is a
danger that so-called `epidemiological evidence' will carry a false air
of authority.''\footnote{\emph{Sienkiewicz v Greif (UK) Ltd} [2011]
  2 AC 229 at p299, paragraph 206, per Lord Kerr of Tonaghmore JSC.} The
courts also recognise, however, that epidemiological evidence used with
proper caution can be admissible and relevant \emph{in conjunction with
specific evidence related to individual circumstances and parties to a
case}. The significance a court may attach to such evidence must depend
on the nature of the epidemiological evidence, and of the particular
factual issues before the court. The point was pithily made by Lord
Rodger of Earlsferry that ``Where there is epidemiological evidence of
association, the court should not proceed to find a causal relationship
without further, non-statistical evidence'',\footnote{Also in
  \emph{Sienkiewicz}, at p287, paragraph 163.} and by Baroness Hale of
Richmond:

\begin{quote}
``The fact that there are twice as many blue as yellow taxis about
on the roads may double the risk that, if I am run over by a taxi, it
will be by a blue rather than a yellow one. It may make it easier to
predict that, if I am run over by a taxi, it will be by a blue rather
than a yellow one. But when I am actually run over it does not prove
that it was a blue taxi rather than a yellow taxi which was responsible.
Likewise, if I actually develop breast cancer, the fact that there is a
statistically significant relationship between, say, age at first
child-bearing and developing the disease does not mean that that is what
caused me to do so.''\footnote{\emph{Sienkiewicz}, p290, paragraph 171.}
\end{quote}

The criteria set out by Bradford Hill for inferring causality in
epidemiological studies are also relevant here.\footnote{Hill, A. B.,
  1965.}

To properly advise their clients, frame the issues for the court, and
present evidence and argument, lawyers need to (that is, should)
understand and effectively deploy the statistical and epidemiological
evidence and the inferences arising from them.

Evidence presented in a case at law can be regarded as data, and the
issue to be decided by the court as a hypothesis under test. The
relationship between these may be immediate, or else indirect, involving
a long chain of intermediate propositions. The outcome of a criminal
trial cannot be known in advance, as it requires sifting, evaluating,
and deciding one or more issues; there is uncertainty about the
relationship of the issue to the evidence such that there is a burden of
proof. Such uncertainty can, in principle at least, be described
probabilistically. We do not suggest that judges and juries are likely
to have (or should be expected to acquire) a sophisticated understanding
of probability or facility in manipulating probabilities; nor that
explicit probability arguments should become routine in courts of
law.\footnote{Although, in particular jurisdictions there may be
  relevant requirements for admissibility, eg in the USA, the Daubert
  standard for evaluating scientific evidence.} There are however
increasing numbers of cases where evidence about probabilities is
clearly relevant, and the court would stand to benefit from advice about
how to handle them.

This section provides advice for lawyers and judges on how to deal with
such cases when they reach the courts. As the previous sections have
discussed, these cases are complex and can be extraordinarily difficult
to investigate. If the investigations lead to prosecution, the evidence
produced in court can be difficult to evaluate. In recent years, serious
concerns have been raised about the fairness of the legal process in a
number of these cases. Here are some suggestions about steps to take and
pitfalls to avoid in order that these cases are tried fairly and justice
is served.

\hypertarget{evaluating-event-clusters}{%
\subsection{Evaluating event clusters}\label{evaluating-event-clusters}}

The first recommendation is to avoid giving undue weight to seemingly
unlikely clusters of events. As discussed in Section 2, seemingly
improbable clusters of events can arise by chance without criminal
behaviour. Consequently, evidence involving event clusters may be less
probative than people assume for distinguishing criminality from
coincidence. Even if it is highly improbable that such a cluster would
occur by coincidence, the best explanation might nevertheless be
coincidence in the absence of convincing evidence to the contrary.
Lawyers and judges should keep this point in mind. In jurisdictions that
rely on lay juries, judicial instructions on this point may be helpful.

In the absence of other evidence, an unlikely cluster of evidence is
difficult to evaluate and may be meaningless. When combined with other
evidence, however, such events can be highly probative. Consequently, it
is extremely important for triers of fact to consider whether evidence
of an event cluster (eg, a surprising number of deaths among the
patients of a particular medical professional) is supported by other
more traditional evidence suggesting that the suspect had the motive,
means and opportunity to kill patients. Failure to find such supportive
evidence, when it would be expected, may constitute strong evidence
against the theory that the suspected individual engaged in
misconduct.\footnote{See commentary of Aart de Vos, translated in Gill,
  2021; Thompson \& Scurich, 2018.} Judicial instructions on this point
might also be helpful in some cases.

\hypertarget{recognising-the-consequences-of-investigative-bias}{%
\subsection{Recognising the consequences of investigative
bias}\label{recognising-the-consequences-of-investigative-bias}}

A second recommendation is to be mindful of the dangers of investigative
bias and the ways that it can unfairly bolster the apparent strength of
the evidence. As discussed in Section 4, bias may arise from
investigators' failure to consider all possible explanations for the
deaths or other negative outcomes under investigation. Consequently, it
is vitally important for lawyers, judges and jurors to consider
carefully whether all potential causal factors have been considered.
That will typically require expert assessments by individuals broadly
knowledgeable about clinical medicine and clinical practice, and often
other fields as well, such as epidemiology and statistics.

If it appears that the initial investigation may have been incomplete
because investigators had insufficient access to independent expert
advice or focused prematurely on an incomplete set of hypotheses, it is
vital that independent experts be called upon to examine the evidence
before trial. Courts should call such experts in jurisdictions where
experts report to the courts, whether independent court appointed
experts or single joint experts. In jurisdictions in which the parties
typically provide courtroom experts, steps should be taken to assure
that such experts are allowed and that parties can afford to provide
them.\footnote{We recognize, of course, that public resources are
  limited and that a balance must be struck between the needs of the
  parties for expert assistance and other priorities. Our goal is to
  explain the importance of expert assistance on this matter, not to
  comment on the priority it should receive relative to other needs.}

Courts should of course consider any alternative explanations offered by
the individual under suspicion. It would be a mistake, however, to
assume that the person under suspicion has sufficient knowledge or
insight to identify all possible causal factors. During the trial of
Dutch nurse Lucia de Berk (discussed in Section 3), who was convicted of
murdering patients but later exonerated, the Judges asked de Berk
whether she could explain why there had been so many deaths among her
patients. They specifically asked her to comment on such matters as
whether she lacked competence, whether her case load was more difficult,
whether she had more night shifts. She could offer nothing to help her
own case. Yet independent experts who examined her case after she was
convicted identified a large number of potential causal factors that
cast the case in an entirely different light, and ultimately contributed
to de Berk's release from prison. The initial investigation had missed
or ignored some of these factors, perhaps because they cast a negative
light on individuals who were involved in the initial investigation. The
Lucia de Berk case is thus an important cautionary tale about the need
to involve independent experts before trial to avoid subsequent
miscarriages of justice. Ideally this will occur during the initial
investigation, but if not, then it needs to be done before the case
comes to trial.

The fairness of the investigation may also be undermined by the failure
of investigators to take adequate steps to mitigate contextual bias. As
illustrated in Section 4, predictable biases may arise when experts
assess such matters as whether a death was ``suspicious'' and whether
the suspect had access to the decedent. Even if the effect of such
biases on the number of deaths counted against the suspect is relatively
small, the cumulative effect can be dramatic on statistics used to
assess the significance of event clusters, such as \emph{p}-values.
Lawyers and judges need to understand that investigative bias can create
seemingly powerful statistical evidence against someone who is entirely
innocent. In light of that insight, they must consider whether adequate
procedures were taken to control bias in the instant investigation and,
if not, how that failure affects the probative value of the statistical
evidence generated by the investigation.

A poorly conducted investigation may yield statistical findings that are
so problematic that they do not warrant consideration. Jurisdictions
that rely on lay juries as triers of fact often require judges to screen
scientific testimony to assure it is sufficiently trustworthy to be
admitted into evidence. In the United States, for example, Rule 702 of
the Federal Rules of Evidence requires the trial judge to determine that
proffered expert testimony ``will help the trier of fact to understand
the evidence,'' that it is ``based on sufficient facts or data,'' that
it is ``the product of reliable principles and methods,'' and that the
expert ``has reliably applied the principles and methods to the facts of
the case.''\footnote{The United States Supreme Court has addressed the
  admissibility of expert evidence under the Federal Rules in a series
  of cases that began with Daubert v. Merrell Dow Pharmaceuticals (1993)
  and included General Electric v. Joiner (1997) and Kumho Tire v.
  Carmichael (1999). All three cases emphasized the judge's role as a
  ``gatekeeper'' with the authority to exclude from the trial expert
  evidence that that is insufficiently ``reliable'' to be trustworthy.}
If the underlying investigation failed to consider relevant causal
factors that might provide an alternative explanation for a cluster of
deaths, and carried out the assessment linking the defendant to those
deaths in a biased manner, it might well be appropriate for a judge to
find that the resulting evidence does not meet the requirements of Rule
702 and should be excluded. Such evidence might also be subject to
challenge under provisions such as Rule 403 of the Federal Rules of
Evidence, which allows a trial judge to exclude evidence from
consideration by a jury ``if its probative value is substantially
outweighed by a danger of \ldots{} unfair prejudice, confusing the
issues, [or] misleading the jury \ldots''

Similar rules apply in England and Wales.\footnote{Kennedy v Cordia
  Services LLP 2016 SC (UKSC) 59, paragraphs [38] \emph{et seq}.;
  see also Forensic Science Regulator (2015, 2021a, 2021b, 2022).} To
assist the court, skilled (or expert) witnesses, unlike other witnesses,
can give evidence of their opinions, if and only if they fall within
their domain of proven and relevant expertise.\footnote{The Professor
  Meadows and shaken-baby syndrome cases, were cautionary examples of an
  expert straying outside their domain of expertise (medicine) and being
  also regarded by the court as expert in their non-expert opinion area
  (statistics).} If on the proven facts a judge or jury can form their
own conclusions without help, then the opinion of an expert is
unnecessary (and thus inadmissible).\footnote{In the case of Ben Geen,
  see Gill RD, Fenton N, \& Lagnado D (2022), the judge ruled that
  written opinions on the biasedness of the investigation submitted by
  two experts in the field of statistics and medicine was merely common
  sense. The experts were not allowed to present their opinions to the
  jury.} As with judicial or other opinions, what should carry weight is
the quality of the expert's reasoning, not whether the expert's
conclusions accord with other evidence. The expert should be careful to
recognise, however, the need to avoid supplanting the court's role as
the ultimate decision-maker on matters that are central to the outcome
of the case. The expert's role is to provide information and analysis
that is helpful to the trier-of-fact, not to comment directly on how the
trier-of-fact should decide the case. On the question of impartiality
and other duties of an expert, these include:

Expert evidence presented to the court should be, and should be seen to
be, the independent product of the expert uninfluenced as to form or
content by the exigencies of litigation.

An expert witness should provide independent assistance to the court by
way of objective unbiased opinion in relation to matters within his or
her expertise. An expert witness in the High Court should never assume
the role of an advocate.

An expert witness should state the facts and assumptions on which the
opinion is based, and should not omit to consider material facts which
could detract from the concluded opinion.

An expert witness should make it clear when a particular question or
issue falls outside his or her expertise.

If an expert's opinion is not properly researched because
insufficient data is available, then this must be stated with an
indication that the opinion is no more than a provisional one. In cases
where an expert witness who has prepared a report could not assert that
the report contained the truth, the whole truth and nothing but the
truth without some qualification, that qualification should be stated in
the report.

If, after exchange of reports, an expert witness changes his or her view
on a material matter having read the other side's expert's
report or for any other reason, such change of view should be
communicated (through legal representatives) to the other side without
delay and when appropriate to the court.

Where expert evidence refers to photographs, plans, calculations,
analyses, measurements, survey reports or other similar documents, these
must be provided to the opposite party at the same time as the exchange
of reports.\footnote{\emph{Kennedy}, p74, para [52], quoting from
  well-established case law.} This applies also to software.\footnote{Similar
  best practice rules for experts are included in both the UK CPR Common
  Procedure Rules and various International Arbitration rules.}

In many jurisdictions, particularly those that use professional judges
as triers of fact, there are no rules for exclusion of untrustworthy or
unreliable scientific evidence. In those jurisdictions statistics
generated in a poorly conducted investigation would need to be
considered but could, of course, be dismissed or ignored if the judges
found them unpersuasive.

While some investigations may be sufficiently problematic to justify
excluding or ignoring the statistical findings entirely, courts are
likely, in most cases, to treat investigative flaws, methodological
limitations and potential biases as issues going to the weight of the
evidence -- that is, as issues for the trier of fact to consider when
weighing the value of the evidence. In light of the issues discussed in
this report, it should be clear that advice, reports and expert
testimony from independent statisticians may be extremely important. If
investigative bias is a significant concern, lawyers and courts should
also consider seeking evaluations from experts of cognitive bias and
factors associated with the accuracy of expert judgment.

\hypertarget{avoiding-fallacious-interpretations-of-statistical-findings}{%
\subsection{Avoiding fallacious interpretations of statistical
findings}\label{avoiding-fallacious-interpretations-of-statistical-findings}}

A third recommendation is to be mindful of the danger of drawing
illogical conclusions from statistical findings, such as
\emph{p}-values, and to take steps to assure that misinterpretation of
statistical findings does not undermine the fairness of the trial. As
discussed in Section 2, people often transpose conditional
probabilities, which can cause them to draw illogical and unwarranted
conclusions from statistics like \emph{p}-values, conclusions that may
be quite unfair to an accused individual.

The first step in avoiding unfairness is for lawyers and judges to
educate themselves about the proper interpretation of such statistics,
so that they can avoid inadvertently incorporating such errors into
arguments they make to the triers of facts or summations of evidence. It
is also important that lawyers and judges avoid eliciting from experts
testimony that incorporates or is conducive to such errors.

Avoiding error in the presentation of evidence is only the first step.
Because people often jump to illogical conclusions on their own, it is
not enough to present the evidence correctly. The trier of fact is
likely to need guidance on correct interpretation of such statistics. If
the triers of fact are professional judges, that guidance could be part
of their professional training. Some jurisdictions are exploring the
possibility of special education in probability for judges who will
handle cases involving statistical evidence; such training would surely
be appropriate for judges handling this class of cases.

With lay triers of fact, the guidance can take two forms. It could be
incorporated into expert testimony. For example, experts could be asked
to comment on the meaning of a \emph{p}-value or similar statistic,
which might allow them to identify incorrect interpretations. An expert
might say, for example that a low \emph{p}-value does not necessarily
imply a low probability that the findings are coincidental. It means
that the evidence observed is unlikely if coincidence is the underlying
explanation, but coincidence may still be more likely than other
explanations in the absence of convincing evidence supporting another
explanation. In jurisdictions where judges instruct the jury on applying
the law to the facts, guidance to this effect might also be incorporated
into judge's instruction.

\section{Conclusions and summary of recommendations}

In this final section, we draw together our main recommendations. We
reiterate that the scope of this report is the use of evidence based on
statistical analysis in cases of suspected medical malpractice. Some of
our recommendations may be appropriate in other contexts, but that is
for others to say. We also recall that our scope is not limited to any
particular jurisdiction; in some jurisdictions some of our
recommendations may be redundant as they advocate what is already
accepted practice.

It should be clear now that in our view, the statistical aspects of
these cases are often nontrivial, fraught with difficulties, challenging
to laypeople (jurors, media reporters, the public) and to lawyers. They
are not entirely straightforward to the specialists!

\medskip

\emph{Recommendation 1:} It is therefore important that all parties
involved in investigation and prosecution in such cases~consult with
professional statisticians, and use only such appropriately qualified
individuals as expert witnesses. [Section 5(c)]

\medskip

There are two kinds of error in drawing inferences about effects from
data: inferring an effect that is not real, or missing one that is. Both
have grave effects in the judicial setting. It has been argued that if
one decreases the error rate of one of the two kinds, the error rate of
the other kind will go up; thus any change in practice shifts the
balance between prosecutor and defence, shifting the errors from Type 1
to Type 2 or vice versa. That is only the case if nothing is changed in
statistical methodology, apart from merely shifting a decision
threshold. But one can reduce both error rates by increasing the amount
of information extracted from the already available data, using superior
statistical methods, and of course by acquiring more and different kinds
of data.

\medskip

\emph{Recommendation 2:} In presenting the results of statistical tests,
both the level of statistical significance (\emph{p}-value) and the
estimated effect size should be stated. One addresses the question of
whether an effect is truly \emph{detected}, the other quantifies the
\emph{size} of that effect, if it exists. These are different concepts
and both are important; neither should be confused with subjective
judgements about the credibility of the expert witness. [Section 4(c),
Section 5, and Appendix 2]

\medskip

Special care is needed to assure that \emph{p}-values, when presented in
reports and testimony, are understood and used properly. While
\emph{p}-values are an important statistical and scientific tool, they
are difficult for people to understand and are frequently
misinterpreted. They may, for example, be misunderstood as statements
about the probability that a coincidence occurred, rather that the
probability of observing a given number of deaths (or more) by chance,
and this kind of misinterpretation can be extremely unfair to
individuals suspected of misconduct.

\medskip

\emph{Recommendation 3:} In reports and testimony, experts should take
care to explain the proper interpretation of p-values and should avoid
drawing fallacious inferences from them. In jurisdictions that rely on
lay jurors, judges should consider providing instructions about the
proper use of \emph{p}-values. Lawyers, judges and investigators should
educate themselves to the dangers of fallacious statistical
interpretation. Lawyers should endeavour to present the case in a manner
conducive to correct understanding, avoiding to the extent possible
testimony or arguments conducive to misinterpretations.

\medskip

We have highlighted the importance of taking a broad and informed view
of all the circumstances in which a cluster of adverse outcomes is
observed, to ensure that all potential causal factors are identified,
and the problem that those best-informed may be implicated in
alternative explanations for the data, with a consequent risk of bias.
We therefore advocated that

\medskip

\emph{Recommendation 4:} Investigations should be guided by panels
representing all relevant areas of expertise but independent of both the
suspect and the employing institution. [Section 5(a)]

\medskip

Statistical investigations of the kind discussed here are not controlled
experiments, but observational studies directed by humans, with all the
inherent unconscious biases pervading all human reasoning. It is
impossible to eliminate completely the role of human judgement in
organising and conducting statistical data acquisition and analysis, but

\medskip

\emph{Recommendation 5:} To the maximum extent practicable, experts
informing an investigation, such as DNA specialists, fingerprint
examiners, toxicologists, and pathologists should be kept ``blind'' to
all aspects of the case irrelevant to the question they are being asked
to answer. Blinding is a key tool in minimising prejudicial subjective
effects such as unconscious bias. [Section 5(b)]

\medskip

Guidelines of this nature for evidence of other kinds already exist in
some jurisdictions. For example, organisations that issue practice
guidelines for matters such as DNA evidence include SWGDAM (USA), FSR
(England and Wales), ENFSI (Europe), and the International society of
Forensic Genetics (ISFG; International).\footnote{In England and Wales,
  the Forensic Science Regulator has issued Codes of Practice and
  Conduct for Forensic Science Providers and Practitioners, with recent
  proposed updates; see Forensic Science Regulator (2021b, 2022).} Our
recommendation for blinding is more comprehensive than what is currently
required in most jurisdictions.

A second universal consequence of basing decisions about causes of
effects on observational studies is captured by the well-known aphorism
``correlation is not causation''.

\medskip

\emph{Recommendation 6:} It is vital that investigators appreciate the
truth of this, and the fact that the connection between them is
well-studied, and that in fields such as medical diagnosis there are
accepted criteria to guide the valid drawing of conclusions in
observational studies [Section 6 and Appendix 7]. Possible
confounding factors must be identified, and their effect quantified,
before attributing causes to observed effects. [Sections 2, 4(a,c)]

\medskip

This report is designed to promote stronger, more scientifically
rigorous investigations of alleged medical misconduct. While that is the
ideal, courts may still occasionally be called upon to evaluate evidence
generated by poorly conducted investigations that produce problematic
results. In jurisdictions that rely on lay juries as triers of fact,
judges should consider whether the results of such an investigation are
sufficiently reliable and trustworthy to meet legal standards for
admissibility.

\medskip

\emph{Recommendation 7:} When courts must evaluate the results of
problematic investigations, it is particularly important that they
consider reports and expert testimony from independent statisticians. If
investigative bias is a significant concern, lawyers and courts should
also consider seeking evaluations from experts of cognitive bias and
factors associated with the accuracy of expert judgment.

\medskip

Understandably, most participants in the legal world have little
training in matters of statistics and the scientific evaluation of
uncertainty. In some countries, organisations in parts of the legal
community ensure that training is available to those who would like it
on probabilistic reasoning, statistical modelling, and statistical
inference. In our opinion, defence lawyers first of all~need to know
that there is a whole scientific field out there which can help them
serve their clients better. They need to be able to learn about the
possibilities and to know how to find the professional community which
can help them. Similarly, prosecution lawyers will need to learn about
these matters -- and if they do not, can expect cases built around
inadequate statistical analysis to be successfully challenged by defence
lawyers with aid of expert testimony. Judges too will need to be
sufficiently informed to be able to determine admissibility and guide
jurors accordingly. Not every legal professional needs to know
everything: obviously, they cannot. However, within the different parts
of the legal community, there do need to be people who do understand
enough to know when professional support and further education is
necessary. Our final, strong recommendation is therefore that

\medskip

\emph{Recommendation 8:} Further interaction between legal and
statistical communities should be fostered by the leaders of the legal
and statistical communities, with a view to promoting joint~educational
activities.

}

\appendix

\section{Appendices}
\subsection{Probability and odds}

The uncertainty of an event can be expressed quantitatively in several
different ways. A probability is a number between 0 and 1, an impossible
event having probability 0 and a certain one probability 1. For
experiments that could be repeated indefinitely, the probability of an
event can be interpreted as the proportion of times the event occurs in
a very large number of independent trials. Another interpretation is the
fair price in £ to buy or sell a bet that pays £1 if the event occurs
and £0 if it does not.

Gamblers usually quantify uncertainty on a different scale, that of
``odds''. In a race between 10 horses of equal ability so that the
result is determined by chance, we would say that the odds on a
particular horse winning were ``9 to 1 against'', equal to a
probability of 1/10 = 0.1. In general the odds for an event of
probability \emph{p} are $(1/p)-1$ to 1 against, and the
probability of an event for which the odds are $O$ to 1 against is
$1/(O+1)$.

In games between two unequal parties, say football teams, the odds
against the weaker team winning might be quoted as 3 to 1, that is a
probability of 1/4 = 0.25. This is the same as saying that the stronger
team has odds of 3 to 1 ``on'' or ``in favour of'' winning
(the possibility of a draw is being neglected here). These are all ways
of suggesting that in a long run of games between the teams, the
stronger team would win 3 times as many as the weaker. Bookmakers have
to make a living, so these numbers are different from those offered when
inviting bets, of course.

There are many different traditions across cultures and countries for
expressing odds verbally or typographically; to avoid ambiguity it is
important to be clear whether the odds quoted are ``against'' or
``in favour of'', and not to simply state, eg, ``the odds are
5''.

\subsection{Statistical significance, effect size and risk}

It is important to be clear about the distinction between absolute and
relative risk, and the role of statistical significance in reporting
changes in risk. These issues are discussed in standard textbooks, so we
just give a brief summary here, and set this summary in the simplest
possible context of a ``before/after'' comparison. Before an
intervention (for example, the appointment of a new nurse), we observe a
certain proportion of adverse outcomes (for example, unexplained deaths
of patients); after that intervention, we see a higher proportion -- how
should we interpret this? We assume here that there are no other changes
in circumstances; in Sections 3 and 4 we discuss at length reasons for
caution in making this assumption.

First, statistical significance -- in contrast to what we might call
scientific significance or importance -- is simply a statement about a
\emph{p}-value. Recall that a \emph{p}-value is the probability that the
data would show a change as large as that observed, or more, under the
assumption (the ``null hypothesis'') that there is no causal effect
of the intervention -- the difference is just due to chance, the nurse
is innocent of wrong-doing. If the \emph{p}-value is very small, we
would be so surprised to see these data that we prefer to disbelieve the
null hypothesis; if it is large, we will conclude that the evidence from
the data is inconclusive, we accept that chance is a plausible
explanation.

To help convey this interpretation of the \emph{p}-value verbally, in
many disciplines we use the idea of \``statistical significance''. A
result is typically termed ``statistically significant'' if the
\emph{p}-value is less than 0.05 (5\%); sometimes for emphasis we write
``statistically significant at the 5\% level''. Commonly, the terms
``highly statistically significant'' or ``very highly
statistically significant'' are used if the \emph{p}-value is less than
0.01 (1\%), or 0.001 (one in a thousand) respectively, but these terms
are not universal. It is usually considered good practice now to simply
report the numerical \emph{p}-value.

As already stated, statistical significance is not the same as
real-world importance, it is simply a statement about whether the
results could be explained by chance. We quantify importance in the idea
of ``effect size''. This is measured in different ways in different
kinds of analysis; for our before/after study we would typically
quantify effect size using relative risk or difference in absolute risk.
If the proportion of adverse events increased from 0.04 to 0.06, this
could be reported as a relative risk of 1.50, or an increase in absolute
risk of 0.02. These two statements mean exactly the same thing, but
there is an evident possibility of their being interpreted differently
by the casual reader, and evident opportunities for sensationalising
changes by choice of presentation -- especially when expressed as
percentages. Is this a 50\% change (in relative risk, from 1.00 to
1.50), or a 2\% change (in absolute risk, from 4\% to 6\%)?

There is no general relationship between statistical significance and
effect size. While it is true that within the context of a single study
on a fixed number of subjects, \emph{p}-values go down as effect sizes
go up, there are no other implications. Table 1 below illustrates this.
Comparing cases (a) and (b) we see that a given effect size in a small
study may not be statistically significant, while the same effect size
in a large study would be, another consequence of the fact that in small
samples the law of large numbers does not dominate, as discussed in
Section 4(f). Comparing (b) with (c) and (d), we see that for rare
events, changes appear greater if expressed using relative risks, and
that for a given relative risk, a rare event is less statistically
significant. Finally, studying (c), we see that even with 800 patients
considered, a doubling of the risk from 5/400 to 10/400 is not
statistically significant.

\begin{center}
\begin{tabular}{|c|c|c|c|}
\hline
Case & \emph{p}-value & Relative risk & Absolute risk difference \\
\hline
a & 0.15 & 2.00 & 0.125 \\
\hline
b & 0.000006 & 2.00 & 0.125 \\
\hline
c & 0.19 & 2.00 & 0.0125 \\
\hline
d & 10\textsuperscript{--10} & 11.00 & 0.125 \\
\hline
\end{tabular}
\end{center}

\begin{longtable}[]{@{}lllllll@{}}
\endhead
& \underline{Case (a)} & & &  & \underline{Case (b)}   & \\
& & & & & & \\
Outcome & Before & After & & Outcome & Before & After \\
Adverse & 5 & 10 & & Adverse & 50 & 100 \\
Normal & 35 & 30 &~~~~ & Normal & 350 & 300 \\
& & & & & & \\
& & & & & & \\
&  \underline{Case (c)}  & & & & \underline{Case (d)}  & \\
& & & & & & \\
Outcome & Before & After & & Outcome & Before & After \\
Adverse & 5 & 10 & & Adverse & 5 & 55 \\
Normal & 395 & 390 & & Normal & 395 & 345 \\
\end{longtable}

\noindent\emph{\textbf{Table 1. Comparing p-values, and absolute and relative
risks for 4 artificial data sets, tabulated above}}

\subsection{Sensitivity and specificity}

In testing a binary, true/false, statement, two kinds of errors can be
made -- to decide true when it should be false, or false when it should
be true. Particularly in medical testing, we use the terms sensitivity
and specificity to quantify how well the test avoids these two kinds of
error. The probability of getting a \underline{positive} test result on
a patient who \underline{does} have the disease is called the
\emph{sensitivity}; the probability of getting a \underline{negative}
test result on a patient who does \underline{not} have the disease is
called the \emph{specificity}. Ideally both should be near to 1 (100\%).
It is not meaningful to speak of the ``accuracy'' of a test: the
proportion of patients whose test give the ``right'' answer depends on
the disease prevalence. Bayes' rule evaluates this probability.

\subsection{Bayes' rule}

To illustrate how to avoid the Prosecutor's Fallacy, and correctly
``invert the conditional'' to calculate the probability of the
hypothesis given the evidence, let us start by considering a different
example in detail. The approach we describe will allow the proper
quantitative comparison of competing explanations for the data at hand.

\medskip

\noindent\emph{\underline{Testing for a disease}}

\noindent Suppose we are trying to determine whether a medical patient has a
particular disease after learning that the patient had a positive test
result for that disease. Let us suppose that two kinds of evidence are
available: evidence about the prevalence of the disease among people
like the patient, and evidence about the accuracy of the diagnostic
test. Assume the test correctly shows positive results for 9 out of 10
people who have the disease, but produces false positive results for 1
person in 100 who does not have the disease. What is the probability
that our patient, who had a positive test result, has the
disease?\footnote{Over the course of the Covid pandemic, most readers
  will have become very familiar with situations like this, and the
  interplay between population prevalence, and sensitivity and
  specificity of different types of test. See also Appendix 3.}

Because there is only 1 chance in 100 that the patient will have a
positive test result \emph{if} the patient does not have the disease, it
might be tempting to assume that there is 1 chance in 100 the patient
does not have the disease and, therefore, 99 chances in 100 that she
does have the disease. By now, however, readers should be sceptical of
this logic. As we have explained, the probability of A given B is not
necessarily the same as the probability of B given A. But what
conclusions can be drawn?

To draw conclusions, we must consider the test result in connection with
other evidence -- in this case, information about the prevalence of the
disease among people like the patient. Let's assume that it is a rare
disease found in only 1 person in 1,000 in people like the patient. With
this additional information, we can draw conclusions, as laid out in the
following paragraph, and depicted in Figure 1 and Table 2.

Consider 1 million people like this patient. If 1 in 1,000 have the
disease, we would expect there to be 1,000 who have it and 999,000 who
do not have it. So the overall chances of having this disease among
people like our patient will be 1 in 1,000. The diagnostic test greatly
shrinks the group without the disease, however, by eliminating the 99\%
of them who give a negative test result. Among the 999,000 who do not
have the disease, only 9,990 will give a positive test result. Because
the test also produces a positive result for 9 out of 10 people who do
have the disease, we would expect 900 of the 1,000 people who have the
disease to give a positive test result, so altogether there will be
9,990 + 900 = 10,890 positive results. But notice that that means there
will be more positive test results among people who do NOT have the
disease (9,990) than among people who do have the disease (900). So the
chances that a person with a positive test result actually has the
disease is only 900/10,890 or about 1 in 12. These calculations are
illustrated in Figure 1 and Table 2.

The positive test result has certainly helped assessment of the patient.
Before knowing the test result, the odds were 999:1 against the patient
having the disease; after learning of the positive test result, the odds
have shortened considerably to about 11.1 against. (The relationship
between probability and odds is elaborated in Appendix 1). The test
result seems highly indicative of disease. Nevertheless, when considered
in light of the other evidence, the odds are still against the patient
having the disease, even after learning of the positive test result.

\bigskip
\noindent\includegraphics[width=4.70in]{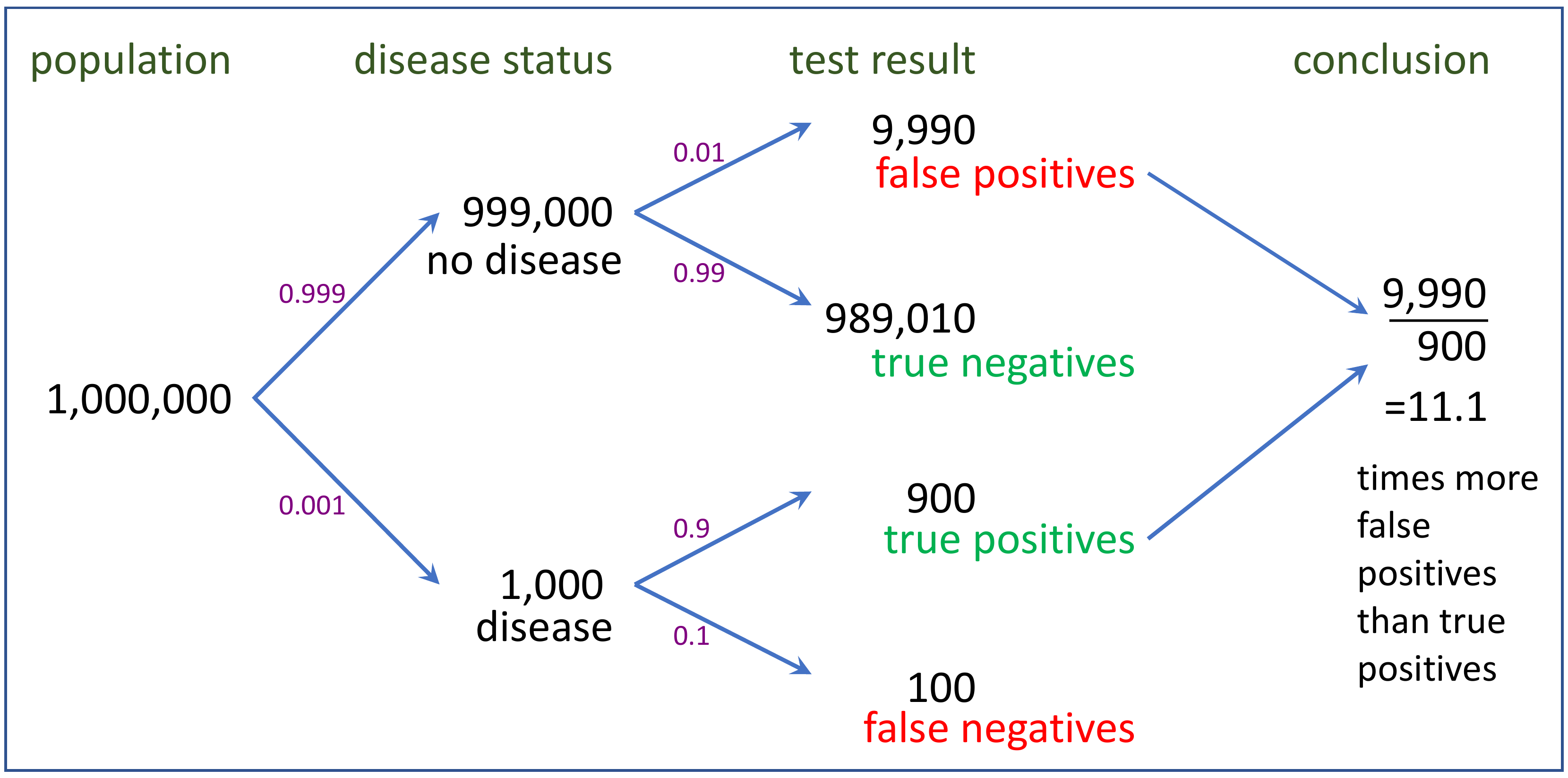}

\medskip

\noindent\emph{\textbf{Figure 1. A simple numerical example of Bayes' rule}}

\bigskip

\begin{longtable}[]{@{}llll@{}}
\toprule
\endhead
& Population & After $+$ve test & After $-$ve test \\
&(prior)&(posterior)&(posterior)\\
\midrule
No Disease & 999,000 & 9,990 & 989,010 \\
Disease & 1,000 & 900 & 100 \\
Total & 1,000,000 & 10,890 & 989,110 \\
Chances of having disease & 1 in 1,000 & 1 in 12.1 & \\
Odds against having disease & 999 to 1 & 11.1 to 1 & \\
\bottomrule
\end{longtable}
\noindent\emph{\textbf{Table 2. A simple numerical example of Bayes' rule}}

\bigskip

The shift in odds described in the previous paragraph can be computed
more quickly by multiplying the prior odds against the disease, 999,000
to 1,000, ie 999 to 1, by what statisticians call the likelihood ratio:
the ratio of the probabilities of a positive result among the ill and
the well, 90\% divided by 1\% ie 90. This approach is quite general, and
is known as Bayes' rule:

\bigskip

\noindent \centerline{POSTERIOR ODDS ~= ~PRIOR ODDS \(\times\) LIKELIHOOD RATIO.}

\bigskip

Later we will explain this reasoning in a more mathematical notation and
in a more general context. Bayes' rule is of course quantitative, but it
is also profitable to understand it qualitatively. On the face of it,
the imperfections in the test do not seem to invalidate it: there is
only a small chance of a false positive, and its usefulness seems nicely
captured by the reassuringly large likelihood ratio, 90. The likelihood
ratio is the ratio of the probabilities of the observed data under the
two hypotheses considered, and is often abbreviated as LR. However, the
test is very poor when used like this in a situation of low disease
prevalence, as we see that a citizen testing positive is over 11 times
more likely to be well than ill. The apparent evidence in favour of the
disease from a positive test result is completely outweighed by the
prior odds that the citizen tested is well. To detect a rare disease
successfully, we need a very much smaller false positive rate (that is,
improved \emph{specificity}).\footnote{As the 18\textsuperscript{th}
  century French mathematician Pierre-Simon Laplace wrote: \emph{We are
  so far from knowing all the agents of nature and their diverse modes
  of action that it would not be philosophical to deny phenomena solely
  because they are inexplicable in the actual state of our knowledge.
  But we ought to examine them with an attention all the more scrupulous
  as it appears more difficult to admit them}. Paraphrased by Carl Sagan
  as \emph{The weight of evidence for an extraordinary claim must be
  proportioned to its strangeness}. See Tressoldi, \emph{Frontiers of
  Psychology}, 2, 117.}
  
\bigskip

\noindent \emph{\underline{Suspected serial homicide}}

\medskip

\noindent Now let us consider how this analysis might apply to a case of suspected
serial homicide by a medical professional. The \emph{p}-value is
analogous to the false positive rate of the diagnostic test -- it tells
us how likely it is to get ``positive evidence'' (a given number of
deaths) if the suspect is not guilty, and the cluster of deaths arose by
coincidence and hence informs us regarding the likelihood ratio. By
itself, the \emph{p}-value cannot tell us the probability the suspect is
guilty or not guilty. But it can provide insight into how much our
assessment of guilt should change in light of the cluster of deaths.
Whether the posterior odds will favour coincidence or misconduct will
depend partly on what other evidence suggests about the ``prior odds''
and partly on the strength of the likelihood ratio.

To summarise, while unexpected clusters of deaths, or other adverse
outcomes, may raise legitimate suspicions and warrant further
investigation, they generally should not be taken as definitive evidence
of misconduct. And this is true even if we accept two key assumptions
underlying the analysis we have just presented, assumptions that may be
problematic in other instances: (1) that the only plausible explanations
for what happened are misconduct by a particular individual and an
extremely unlikely coincidence; and (2) that there is a sound scientific
basis for the probabilities that underlie the LR.

\bigskip

\noindent\emph{\underline{The general case}}

\medskip

\noindent Finally, we give a brief more technical discussion of Bayes' rule.
As an illustration in a case of medical misconduct, let \(H_{p}\) be the
(prosecution) hypothesis that a medical professional engaged in
misconduct that placed his/her patients at an elevated risk of death.
Let \(H_{d}\) be the defence hypothesis that no such misconduct took
place, and hence the risk of death to the patients is the same that
would be expected for other similarly situated patients. Let \(E\) be
the evidence considered in this case -- a specific number of deaths in a
given period in a given ward when the medical professional is on duty.
The adjudicator needs to assess the conditional probability for either
hypothesis, given the evidence: \(P(H_{d}|E)\) and \(P(H_{p}|E)\). It
will not usually be possible to assess these directly, however it will
often be reasonable to assess the probability that the evidence would
have arisen, under each of the competing scenarios: \(P(E|H_{d})\)and
\(P(E|H_{p})\), ie the probability of the evidence \(E\) given the
defence hypothesis and given the prosecution hypothesis. Recall that the
\emph{p}-value is not the same as the probability of the evidence \(E\)
given the defence hypothesis \(P(E|H_{d})\), but it is the probability
of the evidence and more extreme evidence given \(H_{d}\)\emph{.}

What investigators/judges/juries want to know is the probability the
prosecution hypothesis \emph{Hp} is true in light of the evidence.
\emph{Bayes' rule,} a trivial consequence of the definition of
conditional probability, tells us that

$$\frac{P(H_{p}|E)}{P(H_{d}|E)}~ = ~ \frac{P(H_{p})}{P(H_{d})} \times \frac{P(E|H_{p})}{P(E|H_{d})}\eqno(1)$$

The posterior odds for comparing \(H_{p}\) and \(H_{d}\), given the
evidence \(E\) is a simple transformation of \(P(H_{p}|E)\), the desired
posterior probability of \(H_{p}\), \emph{ie} that the suspect engaged
in misconduct. It is important to note that this formulations of Bayes'
rule using odds remains valid whether or not the hypotheses \(H_{p}\)
and \(H_{d}\) are mutually exhaustive; it is not necessary that their
probabilities add to 1. The ``odds'' above are all for one hypothesis
(\(H_{p})\) relative to another (\(H_{d}\)).

The second term on the right-hand side of (1) is constructed out of the
directly assessed terms \(P(E|H_{d})\)and \(P(E|H_{p})\): it is the
likelihood ratio (LR) for \(H_{p}\) as against \(H_{d}\), engendered by
the evidence \(E\). It is noteworthy that only the ratio of these terms
enters, their absolute values being otherwise irrelevant. The first term
right-hand side of (1) \(P(H_{p})\)/\(P(H_{d})\) is the prior odds for
comparing \(H_{p}\) and \(H_{d}\) (\emph{ie}, before the evidence
\emph{E} regarding a high mortality rate is incorporated). The prior
odds might reasonably vary from one individual adjudicator to another.

When \(E\) denotes all the evidence in the case, all the probabilities
in (1) are unconditional; in particular, the prior odds should be
assessed on the basis that there is no evidence to distinguish the
suspect from any other potential suspect -- this can be regarded as one
way of formalising the legal doctrine of ``presumption of
innocence'' (which of course is not the same as an assumption of
innocence). When \(E\) denotes a piece of evidence presented in
mid-process, all the probabilities in (1) must be conditioned on the
evidence previously presented: in particular, the ``prior''
probabilities could themselves have been calculated using (1), as
posterior probabilities based on earlier evidence.

\subsection{Cumulative effects of bias: two worked numerical examples}

To illustrate ways in which investigative bias may distort statistical
evidence emerging from investigations of medical misconduct, let us
consider the following hypothetical examples.

\hypertarget{example-1}{%
\subsubsection{Example 1}\label{example-1}}

First, imagine a hospital ward where 1000 patients are treated per year
and where the annual death rate has historically been 10\%. In a given
year, however, the death rate doubles due to a variety of factors. For
example, the increase in deaths might have been really caused by
\begin{itemize}
\item changes in the patient population that increased the number of sicker,
older patients;
\item changes in staffing that affected the quality of care, such as reduced
staffing levels, loss of better qualified staff members, reduced
training, lower morale;
\item changes in administrative procedures that reduced monitoring, or error
checking;
\item counter-productive changes in treatment regimens, or
\item some combination of such factors.
\end{itemize}

Suppose that hospital administrators, alarmed at the sudden increase in
the death rate, look for an explanation and focus their attention on a
nurse who was hired just before the death rates increased. Perhaps the
nurse has eccentric qualities or manifests eccentric behaviour, such as
laughing or making jokes when a patient dies, drawing colleagues'
attention. To investigate the possibility that this nurse is causing
patients to die, they seek to determine whether the nurse can be linked
to ``suspicious deaths'' that occurred on the ward.

One approach would be to compare the number of ``suspicious deaths''
before and after this nurse was hired. Let's suppose that a fair and
unbiased investigation would classify 10\% of the deaths that occurred
as ``suspicious''-- that is, as deaths that could possibly have been due
to homicide. Assuming the nurse in question is completely innocent, and
murdered no one, the result of the investigation would be 10
``suspicious deaths'' among 1000 patients before the nurse was hired,
and 20 in the year after hiring. This finding looks a little
incriminating, although the increase is not statistically significant at
the 5\% level (see Table 3) even if no other factors were at
play.\footnote{Appendix 8 details how this, and other, calculations of
  \emph{p}-values were performed.} The ``relative risk ratio,'' as
calculated by investigators has doubled after the nurse was hired. But
the nurse had nothing to do with it: the increase was caused entirely by
factors related to patient population, staffing and
supervision ---factors that investigators may fail to consider while
focused on the hypothesis that the causal factor was the nurse.

Now let us suppose that the investigation is not fair and unbiased, but
is instead slanted toward incriminating findings due to the
investigators' unconscious cognitive biases. Suppose, for example, that
the rate at which the experts deem deaths to be suspicious increases
from 10\% to 15\% when they are told that the death occurred while an
alleged serial killer was ``on duty,'' and decreases to 5\% when they
are told the death occurred during the alleged killer was not ``on
duty.'' Under those assumptions, 30 ``suspicious deaths'' would be
reported during the period after the suspected nurse was hired and only
5 in the period before, for a ``relative risk ratio'' of 6 (see Table
3). In other words, it now appears that suspicious deaths were six times
more likely after the nurse joined the hospital staff than before. The
\emph{p}-value, indicating the probability of this difference occurring
by chance is about 0.0068 (about 1 chance in 150), which sounds strongly
incriminating for the nurse. But of course, the nurse is entirely
innocent. The seemingly incriminating finding was generated by biased
investigators who failed to take account of other factors that might
have affected the death rate and interpreted the data in a manner that
was inadvertently influenced by predictable human biases. An unbiased
investigation would have shown a smaller (and therefore less
incriminating) increase in deaths.

\bigskip

%\begin{longtable}[]{@{}lllll@{}}
\begin{center}
\begin{tabular}{|c|c|c|c|c|}
\hline
& Patients & Deaths & Deaths deemed & Deaths deemed  \\
& &  & suspicious in& suspicious in\\
& &  & a biased& an unbiased \\
& &  &  investigation & investigation \\
\hline
Before & 1,000 & 100 & 5 & 10 \\
\hline
After & 1,000 & 200 & 30 & 20 \\
\hline
Relative risk ratio & & & 6 & 2 \\
\hline
\emph{p}-value & & & 0.0068 & 0.5876 \\
\hline
\end{tabular}
\end{center}
\noindent\emph{\textbf{Table 3. Number of ``Suspicious Deaths'' Before and After
Suspect Joined the Hospital Staff (as Reported by a Biased
Investigation)}}

\hypertarget{example-2}{%
\subsubsection{Example 2}\label{example-2}}

Another approach that investigators may take is to compare the number of
``suspicious deaths'' when the nurse was or was not on duty. Our second
example illustrates how statistics produced by such comparisons can be
distorted by (a) failure to take account of other causal factors that
may correlate with the duty periods; and (b) investigative bias in
determining which deaths are suspicious. It also allows the time periods
over which the data are collected to be unequal in length.

Suppose that 16 patients die in circumstances assessed by investigators
to be suspicious over a 15-day period on the ward in question, with 9 of
those deaths reported during the 7-hour morning shifts and the remaining
7 during the afternoon and night shifts. The nurse under suspicion works
8 morning shifts, and 2 of the afternoon or night shifts. So the raw
rate of suspicious deaths tends to be higher when the nurse is on duty
than when not, simply by virtue of the nurse's pattern of work. The
first columns in Table 4 (under `Unbiased investigation') tabulate these
values. Compared to Example 1 it is now more difficult to interpret the
data intuitively, but cross-classifying the deaths by shift and the
nurse's presence in this way suggests that the time of day is an
important factor; an appropriate formal method of analysis is described
in Section 5(c) below, and yields the \emph{p}-values in the final line
of the table. These show that allowing for the inherent differences
between shifts transforms the strength of evidence against the nurse
from statistically significant (\emph{p}=0.017) to very weak indeed
(\emph{p}=0.378).

\bigskip

\begin{longtable}[]{@{}llllll@{}}
\toprule
\endhead
&&\multicolumn{4}{c}{Deaths attributed to nurse on duty}\\
\midrule
& & \multicolumn{2}{c}{Unbiased investigation}& \multicolumn{2}{c}{Biased investigation}\\
\midrule
& Shifts & Ignoring& Allowing & Ignoring & Allowing\\
&&morning &morning& morning & morning\\
&&effect &effect &effect &effect\\
\midrule
On duty,  & 8 & 10 & 7 & 12 & 8 \\
morning\\
\midrule
On duty, & 7 & & 3 & & 4 \\
other\\
\midrule
Off duty, & 2 & 6 & 2 & 4 & 1 \\
morning \\
\midrule
Off duty, & 28 & & 4 & & 3 \\
other\\
\midrule
\emph{p}-value for&& 0.017 & 0.378 & 0.0007 & 0.031  \\
nurse effect \\
\bottomrule
\end{longtable}
\noindent\emph{\textbf{Table 4. Numbers of ``suspicious deaths'' when suspect was
and was not on duty, under assumptions of both unbiased and biased
investigations (see text)}}
\bigskip

Finally, let us suppose that cognitive bias also influences the
investigators' assessments of whether each of the deaths was suspicious,
in such a way that 2 additional deaths during the nurse's shifts are now
judged suspicious, one in the morning and one in the afternoon, while
one fewer death was called suspicious in each of the counts where the
nurse was not on duty. The final columns of Table 4 (under `biased
investigation') show these data, and the corresponding \emph{p}-values,
show that this small bias (which might also have been caused by simple
mis-recording in duty records) is enough to make the evidence now
statistically significant (\emph{p}=0.031) even when we assume there is
no difference between morning and other shifts in mean rates of death,
whilst if we do not allow for such differences the evidence is very
highly significant (\emph{p}=0.0007).

All of the analyses here assume that there are no other causal effects,
such as seasonal factors or administrative changes, that need to be
taken into account.

In Appendix 6, we describe analyses of the data in these two examples,
and explain the logic and the calculations that lead to the
\emph{p}-values quoted above.

\subsection{Patterns of occurrence of adverse events}

Here we give some annotated examples of correct analyses of illustrative
data on patterns of occurrence of adverse events. To simplify exposition
we will write about unexpected deaths of patients in a section of a
hospital, and take the ``explanation'' under consideration to be
deliberate harm caused by a nurse. Of course, this exposition applies
\emph{mutatis mutandis} to many other scenarios, and any professional
role, etc.

We will assume that these events occur completely at random, but at a
rate per unit of time (hour, shift, day, etc., as appropriate) that
varies with time, and may be influenced by factors of the kind already
discussed: seasonal and diurnal effects of disease, administrative
changes, etc., and also, possibly, by wilful harm. We use the phrase
``completely at random'' in its proper mathematical sense, to mean that
the occurrence of an event at a particular time has no direct influence
on the time of any other event. (In other words, we consider only
\emph{exogenous} causes for the variation in rate of the adverse event,
not \emph{endogenous} ones). This rules out for example infections of a
contagious disease, where there can be a direct causal link, but would
cover heart attacks. The only other assumption we make is that when we
consider two or more causal factors for the variation in rate, the
effects of these are multiplicative: the percentage change in rate when
one factor is present is the same whether or not other factors are also
present.

As an artificially simple example, consider a hospital ward which is
staffed either by nurse A or by nurse B. Numbers of deaths when each of
the nurses is in charge are counted, and summarised here:

\bigskip

\begin{longtable}[]{@{}llll@{}}
\toprule
\endhead
& Nurse A & Nurse B & Total \\
Died & 15 & 9 & 24 \\
Survived & 25 & 31 & 56 \\
Total & 40 & 40 & 80 \\
\bottomrule
\end{longtable}
\noindent\emph{\textbf{Table 5. Illustrative example: patient survival statistics
under the care of two nurses}}

\medskip

Could the apparent discrepancy in rates of death be attributed to
chance, ``just a coincidence''? We suppose that all circumstances of the
Nurse A and Nurse B shifts are identical; there is no other conceivable
reason for the apparent difference other than the presence of one nurse
or the other.

This data structure is called a (2 by 2) contingency table: the standard
way to analyse this, to test the hypothesis that there is no difference
in the death rates attributable to the nurses, is ``Pearson's
chi-squared test''. This is an elementary technique, taught in the
middle years of high school (eg GCSE level in England and Wales). This
reveals that the probability of observing a difference in apparent death
rates as large as, or larger than, that seen in the table\emph{, if
there was really no difference} is 14\% (that is the \emph{p}-value is
0.14). That means that if you were to repeatedly allocate 24 deaths and
56 survivals into two groups of 40 patients at random, a difference in
apparent rates as large as that in Table 5 would be obtained about 1
time in 7. We have to conclude \emph{there is no significant
difference}. It would be misleading to the court to testify that there
\emph{was} a difference. A \emph{p}-value less than 0.05 is a
pre-requisite for publication in the scientific literature (and this is
not a tough standard, very many scientific ``findings'' are never
replicated by other scientists).

The \emph{p}-value above is calculated as follows. Since 24 out of the
80 patients die, if there were no nurse differences, you would expect
that (24/80) of 40, ie 12, of the deaths would occur on Nurse A's shift.
In the same way, for each of the other counts in the table (9, 25 and
31) you would expect respectively (12, 28 and 28). We denote the
observed numbers (15, 9, 25, 31) by \(O_{i}\) and the expected numbers
(12, 12, 28, 28) by \(E_{i}\), then calculate

\[G = \sum_{i = 1}^{4}\frac{\left( O_{i} - E_{i} \right)^{2}}{E_{i}},\]

(that is, we take each of the cells of the table in turn and square the
difference between the observed and expected numbers, and divide by the
expected number; these fractions are added up over the four cells),
which is 2.143. To convert this to the \emph{p}-value quoted, we can use
standard printed tables of the chi-squared distribution, or the function
found on many calculators and all statistical software packages.

In contrast, if all of the numbers in Table 5 were exactly 10 times
larger (150, 90, and so on), then the Pearson chi-squared statistic
\(G\) would be 21.43 and the \emph{p}-value turns out to be 0.000004, so
there would be overwhelming evidence that the apparent different in
death rate was \emph{not} due to chance. (This is an example of the
point made in Section 4(f) that ``coincidental fluctuations from
population means are more likely with small samples\ldots'').\footnote{See
  also Appendix 2.}

Contingency tables of any size, not just 2 by 2, can be analysed with
Pearson's chi-squared test, but still very few criminal cases are simple
enough to fit into this setting. Nevertheless, the analysis can be
extended to deal with much more complex situations, allowing in
particular more than one causal factor, and different durations of time.
The more general framework is that of \emph{Poisson log-linear models},
which are an example of \emph{generalised linear models}. This is also a
standard methodology, but one now taught not at high school but in
undergraduate courses in mathematics and statistics. When applied to a
2-way contingency table, the results are the same.

These methods are provided in standard statistics packages, and will be
part of the toolbox of all practicing professional statisticians. The
assumptions underlying their use are simply those mentioned above, and
courts should be able to accept results of such analyses in expert
witness testimony, just as, for example, a scientist would expect to be
able to present scientific evidence relying on data from electron
microscopes or mass-spectrometers without needing to explain to judge
and jury the physics needed to say how these complex machines function.
In short, an expert witness, including a statistician, must be free to
use adequate methodology for the task. In Appendix 8, we show computer
code and output for the analyses in this section, using the
well-regarded statistical system R, which is freely and universally
available.

To illustrate appropriate methodology for analysing data on counts of
deaths in different periods in the presence of other possible causal
factors, consider the artificial example from Table 4 of Section 4. The
deaths have been tabulated and summarised in the counts in four
different categories of shifts. Note that these categories differ in
various ways -- they are of different durations; some are morning
shifts, not all; and for some but not all the nurse in question is on
duty. The rates of death vary between the extremes of 4 in 28 shifts and
2 in 2 shifts, a considerable difference, but can we attribute these
differences in rate to the morning/other shift factor, or to the
presence of the nurse, whilst allowing for the fact that among these
small counts there will also be random variation?

To correctly assess the extent to which the deaths can be attributed to
the presence of the nurse we must compare two hypotheses:

\begin{quote}
(1) that the \emph{only} cause of systematic difference in rates is the
shift effect, and

(2) that \emph{both} the shift effect and the presence of the nurse have
a systematic effect on the rates.
\end{quote}

This way of posing the question accords with both sound scientific
practice, and the criminal law principle of \emph{in dubio pro
reo}\footnote{The principle that a defendant may not be convicted by a
  court when doubts about his or her guilt remain.}. It is incorrect,
and pre\-judicial, simply to examine whether the presence of the nurse
affects the rates whilst ignoring the other potential causal factor.

This point is illustrated in the analyses of the Table 4 ``unbiased
investigation'' data summarised in the final rows of that Table and in
Table 6. If we ignore the shift effect, we are simply comparing the
rates of 10 per 15 shifts with 6 per 30 shifts when the nurse is or is
not on duty. The Poisson log-linear analysis (details explained in
Appendix 6, with code in Appendix 8) gives a \emph{p}-value of 1.7\%
(0.017) for the likelihood ratio test that the nurse's presence has no
effect on the rates -- and we would conventionally call this result
significant, which would be incriminating. However, if we follow the
correct practice of comparing hypotheses (1) and (2) above, the
\emph{p}-value becomes 37.8\% (0.378). Because this higher p-value is
not statistically significant, it provides no basis for rejecting
hypothesis (1) and therefore cannot be incriminating. Table 6 also gives
the expected numbers of deaths in each category of shifts, the maximum
likelihood estimates according to the statistical models being fitted in
the two approaches. It is easily verified by inspection that the values
in the case of the correct analysis shown in the 6\textsuperscript{th}
column fit the observed data (4\textsuperscript{th} column) much better
than do the expected numbers under the incorrect analysis
(5\textsuperscript{th} column).

\bigskip

\begin{longtable}[]{@{}llllll@{}}
\toprule
\endhead
Number& Nurse & Shift & Deaths & Expected deaths & Expected deaths \\
of shifts &  & &  &  ignoring  &  allowing  \\
&  & & & morning effect &  morning effect \\
\midrule
8 & on duty & morning & 7 & 5.33 & 7.87 \\
7 & on duty & other & 3 & 4.67 & 2.13 \\
2 & off duty & morning & 2 & 0.40 & 1.13 \\
28 & off duty & other & 4 & 5.60 & 4.87 \\
\bottomrule
\end{longtable}
\noindent\emph{\textbf{Table 6. Continuing the example in Table 4}}

\subsection{Usual practice in medical statistics and epidemiology.}

Two types of clusters of events are routinely investigated in medical
statistics and epidemiology: outbreaks of food poisoning and clusters of
severe adverse events or usually rare diseases. The difficulty of
identifying possible causal factors is multi-faceted. The methods
developed in medicine to bring together evidence from laboratory
science, observational and experimental studies are important tools for
investigation.\footnote{ICCA \& RSS, 2019, Statistics and probability
  for advocates, p.18.}

The standard first approach is to design and conduct a case-control
study, a rapid and relatively inexpensive method. For each precisely
defined incident, one or more control incidents are found, and the
antecedents investigated. The design stage establishes clear definitions
of events, and consistent approaches to seeking evidence for cases and
controls. The varieties of biases which can arise are well-studied, and
methods to minimise the risks of such biases established. It is standard
for those recording data on possible explanatory variables to `be
blind' to which people are cases or controls. For deaths, experts
in the quality and coding of death certificates might provide a
necessary complement to physicians or pathologists.

In the study of causes of disease, nine aspects of association, the
Bradford Hill guidelines, are considered.\footnote{Hill, AB, 1965.} It
would be sensible to consider these in other investigations of causes,
as is happening in areas of civil litigation.\footnote{ICCA \& RSS
  guide, 2019, p.19.} Detailed consideration of uncertainty is
preferable to false confidence in a single explanation.

Comparisons of different authorities' methods of investigating
clusters of events might well lead to mutual benefit.\footnote{Stewart,
  Ghebrehewet \& Jarvis (2016).} The methods in Public Health England
guidelines for investigating non-infectious disease clusters possibly
due to environmental exposures are relevant to clusters of
death.\footnote{Public Health England (2019).} As well as suggested
membership, with roles and responsibilities, of an investigation team,
the guidelines include a substantial list of useful data sources.

\textbf{An example of an efficient investigation} is that of a cluster
of serious events in children with cystic fibrosis.

In 1993, doctors at Alder Hey Children's Hospital (AHCH),
Liverpool, noticed that five children with cystic fibrosis (a condition
in which the lungs and digestive system are clogged with thick sticky
mucus) who needed surgery because of fibrosing colonopathy (obstruction
of the intestine) presented between July and September, 1993. One
response to this might have been to suggest that doctors at AHCH were
failing in some way.

On 8 January 1994, a short report was published, which reported that
``The only consistent change in management had occurred 12--15
months preciously when all five had switched to'' high-strength
pancreatic enzymes (high dose drugs). \footnote{Smyth, et al, 1994.}

At the time the report was published, a case-control study to
investigate the findings had been started: this is the appropriate
method of reacting to the reports of new adverse events among patients.
The Medicines Control Agency had been informed of the cases, and had
issued appropriate warnings. There were about 7600 people known to have
cystic fibrosis in the UK; 5/7600 is 0.07\%.

On 11 November 1995, about 2 years later, the results of the
case-control study were published.\footnote{Smyth, et al, 1995.}

The study had 14 cases of fibrosing colonopathy, with each case matched
to four controls. Data on these 70 patients showed a significant (at
5\%) odds ratio of 1.45 per extra 1,000 high-strength capsules, and
indicated which two particular proprietary formulations were associated
with the highest odds ratios. That is, this association between
particular formulations and fibrosing colonopathy could have arisen by
chance one time in twenty.

Laxative use was also found to be associated with fibrosing colonopathy;
odds ratio 2.42 (95\% confidence interval 1.20--4.94). From a case-control study,
one cannot establish whether laxative use was a cause of fibrosing
colonopathy, or a symptom of it.

Six of the 14 cases received care at AHCH. Care at Liverpool was
associated with approximately a two-fold increase in risk of fibrosing
colonopathy. If taken alone, this risk is statistically significant at
the 4 percent level ($p=0.04\%$), but adjusting for high-dose drugs
removes the significance ($p=0.3$ or $p=0.8$).

In deciding whether to suggest that AHCH doctors were negligent, or
actively harming children with cystic fibrosis, one must consider the
competing explanations for fibrosing colonopathy.

\subsection{Annotated code and output}

This appendix may be of limited interest to readers who are not
statisticians, but is included for two main reasons. One, in the
interests of full disclosure, is to verify the results in the
illustrative numerical examples in Sections 4(f) and 5(c), and the
calculations in Appendix 2, and make them completely reproducible. The
other is that the codes may serve as templates for investigators and
expert witnesses undertaking similar analyses in future, and a starting
point for the more elaborate analyses that will be necessary in many
real cases. These might entail additional explanatory factors,
interactions between them, and possibly additional variables modelled as
random effects. The last-mentioned here will necessitate use of
generalised linear mixed models, available in R by using the lme4
package, for example.

\bigskip

\noindent\emph{\textbf{Analysis in Table 3}}

\bigskip

\noindent\underline{Input}

\bigskip

\noindent Create a 2 by 2 matrix containing the data for the biased investigation:
deaths and survivals for the before and after cases, and display the
data.

{\tt
\begin{verbatim}
biased<-matrix(c(5,95,30,170),2,2)
biased
#
# Conduct Fisher's test for equality of the odds ratios before and after,
# against the alternative that the odds on survival is less.
#
fisher.test(biased,alternative='less')
#
# Repeat for the unbiased investigation
#
unbiased<-matrix(c(10,20,90,180),2,2)
unbiased
fisher.test(unbiased,alternative='less')
\end{verbatim}
}

\bigskip

\noindent\underline{Output}

\bigskip

\noindent Biased case

\medskip

{\tt
\begin{verbatim}
>  biased
[,1] [,2]
[1,] 5 30
[2,] 95 170
> fisher.test(biased,alternative='less')
Fisher's Exact Test for Count Data
data: biased
p-value = 0.006818
alternative hypothesis: true odds ratio is less than 1
95 percent confidence interval:
0.0000000 0.7151649
sample estimates:
odds ratio
0.2992371
\end{verbatim}
}

\bigskip

\noindent Unbiased case

\medskip

{\tt
\begin{verbatim}
> unbiased <- matrix(c(10,20,90,180),2,2)
> unbiased
[,1] [,2]
[1,] 10 90
[2,] 20 180
> fisher.test(unbiased,alternative='less')
Fisher's Exact Test for Count Data
data: unbiased
p-value = 0.5876
alternative hypothesis: true odds ratio is less than 1
95 percent confidence interval:
0.00000 2.08157
sample estimates:
odds ratio
1
\end{verbatim}
}

\bigskip

\noindent\emph{\textbf{Analysis in Tables 4 and 6}}

\bigskip

\noindent\underline{Input}

\medskip

{\tt
\begin{verbatim}
# Create data frame containing the response variables deaths, two
# explanatory factors nurse and morning, and the variable shifts, the
# number of shifts for that row of the table, used on a log-scale as an
# offset since we are modelling rates of deaths per unit time.
shifts<-c(8,7,2,28)
nurse<-as.factor(c('yes','yes','no','no'))
morning<-as.factor(c('yes','no','yes','no'))
deaths<-c(7,3,2,4)
data<-data.frame(shifts,morning,nurse,deaths)
print(data)
# Fit Poisson log-linear models for rates of death, both with just nurse
# included as an explanatory variable, and with morning also included.
# Print analysis of deviance table and fitted values in each case
fitN<-glm(deaths~nurse+offset(log(shifts)),
family=poisson(),data)
print(anova(fitN,test='Chisq'))
print(fitted(fitN))
fitMN<-glm(deaths~morning+nurse+offset(log(shifts)),
family=poisson(),data)
print(anova(fitMN,test='Chisq'))
print(fitted(fitMN))
\end{verbatim}
}

\bigskip

\noindent\underline{Output}

\bigskip

\noindent Display of data frame.
{\tt
\begin{verbatim}
  shifts morning nurse deaths
1      8     yes   yes      7
2      7      no   yes      3
3      2     yes    no      2
4     28      no    no      4
\end{verbatim}
}
\bigskip
\noindent Analysis of deviance table where only nurse is fitted. Note that the
\emph{p}-value for the nurse effect is 0.01728, ie 1.7\%, so apparently
statistically significant.
{\tt
\begin{verbatim}
Analysis of Deviance Table

Model: poisson, link: log

Response: deaths

Terms added sequentially (first to last)


      Df Deviance Resid. Df Resid. Dev Pr(>Chi)  
NULL                      3     10.570           
nurse  1   5.6678         2      4.902  0.01728 *
---
Signif. codes:  0 ‘***’ 0.001 ‘**’ 0.01 ‘*’ 0.05 ‘.’ 0.1 ‘ ’ 1
\end{verbatim}
}
\medskip
\noindent 
Fitted values for this model.
{\tt
\begin{verbatim}
       1        2        3        4 
5.333333 4.666667 0.400000 5.600000 
\end{verbatim}
}
\medskip
\noindent Analysis of deviance table where morning and nurse are both fitted. Note
that the \emph{p}-value for the nurse effect is now 0.37849, i.e.
37.8\%, so is not statistically significant.

{\tt
\begin{verbatim}
Analysis of Deviance Table
Analysis of Deviance Table

Model: poisson, link: log

Response: deaths

Terms added sequentially (first to last)


        Df Deviance Resid. Df Resid. Dev Pr(>Chi)   
NULL                        3    10.5699            
morning  1   8.6617         2     1.9081  0.00325 **
nurse    1   0.7756         1     1.1325  0.37849   
\end{verbatim}
}
\medskip
\noindent 
Fitted values for this model.
{\tt
\begin{verbatim}
       1        2        3        4 
7.872829 2.127171 1.127171 4.872829 
\end{verbatim}
}

\bigskip

\noindent\underline{Biased investigation}

\bigskip

\noindent This proceeds in exactly the same way, but using the biased data
{\tt
\begin{verbatim}
deaths<-c(8,4,1,3)
\end{verbatim}
}
\noindent in the input.

\bigskip

\noindent \emph{\textbf{Analysis in Table 5}}

\bigskip

\noindent\underline{Input}

\bigskip

{\tt
\begin{verbatim}
# Create data frame consisting of a numerical response variable count and
# two factors nurse, the explanatory variable, and died, the response
# category.

nurse<-as.factor(c('A','B','A','B'))
died<-as.factor(c('yes','yes','no','no'))
count<-c(15,9,25,31)
data<-data.frame(died,nurse,count)
print(data)

# Fit a Poisson log-linear model, allowing for main effects nurse and
# died, and an interaction between them.

fit<-glm(count~died*nurse,data,family=poisson())

# Output analysis of deviance table: the interaction term quantifies the
# differential effect of the two nurses on survival.

print(anova(fit,test='Chisq'))

# The analysis of deviance table by convention uses the deviance as the
# test statistic: the following calculation demonstrates that it is
# numerically very similar to Pearson's chi-squared statistic, as defined
# in the text.

E<-c(12,12,28,28)
print(sum((count-E)^2/E))
print(2*sum(count*log(count/E)))
\end{verbatim}
}

\bigskip

\noindent \underline{Output}

\medskip

\noindent Display of data frame.
{\tt
\begin{verbatim}
  died nurse count
1  yes     A    15
2  yes     B     9
3   no     A    25
4   no     B    31
\end{verbatim}
}
\medskip
\noindent
Analysis of deviance table. Note that the \emph{p}-value for the
died:nurse interaction is 0.1416, ie 14.2\%, so not statistically
significant.
\medskip
{\tt
\begin{verbatim}
Analysis of Deviance Table

Model: poisson, link: log

Response: count

Terms added sequentially (first to last)


           Df Deviance Resid. Df Resid. Dev  Pr(>Chi)    
NULL                           3    15.3254              
died        1  13.1653         2     2.1601 0.0002852 ***
nurse       1   0.0000         1     2.1601 1.0000000    
died:nurse  1   2.1601         0     0.0000 0.1416334    
---
Signif. codes:  0 ‘***’ 0.001 ‘**’ 0.01 ‘*’ 0.05 ‘.’ 0.1 ‘ ’ 1
\end{verbatim}
}
\medskip

\noindent Values of the two test statistics, the Pearson chi-squared statistic and
the deviance statistic as used in the table above. They are very similar
numerically, so it is immaterial which is used in calculating the
\emph{p}-value.
\medskip
{\tt
\begin{verbatim}
[1] 2.142857
[1] 2.160122
\end{verbatim}
}

\bigskip

\noindent\emph{\textbf{Calculations in Appendix 2}}

\bigskip

\noindent\underline{Input}

\bigskip

{\tt
\begin{verbatim}
# Set up 4 illustrative data sets
casea<-matrix(c(10,30,5,35),2,2)
caseb<-matrix(c(100,300,50,350),2,2)
casec<-matrix(c(10,390,5,395),2,2)
cased<-matrix(c(55,345,5,395),2,2)
#
# Define function to conduct chi-squared test, and calculate relative risk
# and absolute risk difference
comparerisks<-function(y)
{
ct<-chisq.test(y,,FALSE)
cat('statistic',ct$statistic,' p-value',ct$p.value,'\n')
risks<-y[1,]/apply(y,2,sum);
cat('risks',risks,'\n')
rr<-risks[1]/risks[2]
ard<-risks[1]-risks[2]
cat('RR',rr,' AR difference',ard,'\n')
}
#
# Apply function to data sets
casea
comparerisks(casea)
caseb
comparerisks(caseb)
casec
comparerisks(casec)
cased
comparerisks(cased)
\end{verbatim}
}

\bigskip

\noindent\underline{Output}

\bigskip

{\tt
\begin{verbatim}
> casea
[,1] [,2]
[1,] 10 5
[2,] 30 35

> comparerisks(casea)
statistic 2.051282 p-value 0.1520781
risks 0.25 0.125
RR 2 AR difference 0.125

> caseb
[,1] [,2]
[1,] 100 50
[2,] 300 350

> comparerisks(caseb)
statistic 20.51282 p-value 5.923318e-06
risks 0.25 0.125
RR 2 AR difference 0.125

> case
[,1] [,2]
[1,] 10 5
[2,] 390 395

> comparerisks(casec)
statistic 1.698514 p-value 0.1924825
risks 0.025 0.0125
RR 2 AR difference 0.0125

> cased
[,1] [,2]
[1,] 55 5
[2,] 345 395

> comparerisks(cased)
statistic 45.04505 p-value 1.925539e-11
risks 0.1375 0.0125
RR 11 AR difference 0.125
\end{verbatim}
}

\newpage
\subsection{Members of the working party drawing up this report}

Professor Peter Green FRS, Emeritus Professor of Statistics, University
of Bristol, and Distinguished Professor, University of Technology,
Sydney.

\medskip\noindent Professor Richard Gill, Emeritus Professor of Statistics, Leiden
University.

\medskip\noindent Neil Mackenzie QC, Arnot Manderson Advocates, Edinburgh.

\medskip\noindent Professor Julia Mortera, Professor of Statistics, Università Roma Tre.

\medskip\noindent professor William Thompson, Professor Emeritus of Criminology, Law, and
Society; Psychology and Social Behavior; and Law, University of
California, Irvine.

\medskip\noindent In addition, we are grateful to Professor Jane Hutton, Professor of
Medical Statistics, University of Warwick, for providing Appendix 7.

\bigskip\noindent The report as published by the Royal Statistical Society is available  on the society's webpages at 
\href{https://rss.org.uk/RSS/media/File-library/News/2022/Report_Healthcare_serial_killer_or_coincidence_statistical_issues_in_investigation_of_suspected_medical_misconduct_Sept_2022_FINAL.pdf}{``Full report''} (64~pp.)
A summary version is available at
\href{https://rss.org.uk/RSS/media/File-library/News/2022/Summary_Healthcare_serial_killer_or_coincidence_statistical_issues_in_investigation_of_suspected_medical_misconduct_September_2022_FINAL.pdf}{``Report summary''} (16~pp.)

\end{document}